\documentclass[aps,prd,amsmath,11pt,superscriptaddress,floatfix,nofootinbib]{revtex4-1}
\usepackage{epsfig,amssymb,amsfonts,amsmath,mathtools,bm,color,xcolor,graphicx,braket,adjustbox,esint,upgreek}
\usepackage{multirow}
\usepackage[utf8]{inputenc}
\usepackage{setspace}
\usepackage{dcolumn,kantlipsum}
\allowdisplaybreaks

\usepackage{natbib}
\usepackage{graphicx}

\newcommand{\be}{\begin{equation}}
\newcommand{\ee}{\end{equation}}
\newcommand{\bea}{\begin{eqnarray}}
\newcommand{\eea}{\end{eqnarray}}
\newcommand{\beas}{\begin{eqnarray*}}
\newcommand{\eeas}{\end{eqnarray*}}

\newcommand{\gev}{\mbox{GeV}}
\newcommand{\tev}{\mbox{TeV}}

\def\vec#1{\boldsymbol{#1}}

\newcommand{\scnu}{\affiliation{{Guangdong Provincial Key Laboratory of Nuclear Science,}\\ Institute of Quantum Matter, South China Normal University, Guangzhou 510006, China}}

\newcommand{\snsc}{\affiliation{Guangdong-Hong Kong Joint Laboratory of Quantum Matter, Southern Nuclear Science Computing Center, South China Normal University, Guangzhou 510006, China}}

\newcommand{\ihep}{\affiliation{Institute of High Energy Physics, Chinese Academy of Sciences, Beijing 100049, China}}

\newcommand{\valencia}{\affiliation{Instituto de F\'isica Corpuscular (centro mixto CSIC-UV),
Institutos de Investigaci\'on de Paterna, Apartado 22085, 46071, Valencia, Spain
}}

\begin{document}

\title{Prompt production of the hidden charm pentaquarks in the LHC}

\author{Pan Ling}
\scnu\snsc

\author{Xiao-Hu Dai}
\scnu\snsc

\author{Meng-Lin Du}\email{du.menglin@ific.uv.es}
\valencia

\author{Qian Wang}\email{qianwang@m.scnu.edu.cn}
\scnu\snsc\ihep

\begin{abstract}
Motivated by the observation of the first hidden charm pentaquarks
by the LHCb collaboration in 2015 and the updated analysis with 
an order-of-magnitude larger data set in 2019, we estimate their cross sections for 
the prompt production as well as their heavy quark spin partners,
in the $\Sigma_c^{(*)}\bar{D}^{(*)}$ hadronic molecular picture, 
at the center-of-mass energy $7~\mathrm{TeV}$ in the $pp$ collision. 
Their cross sections are several $\mathrm{nb}$ and we would expect
several tens hidden charm pentaquark events in the LHC based on its current integrated luminosity.
The cross sections show a sizable deviation of the cross sections for hidden charm pentaquarks with
the third isospin component $I_z=+\frac{1}{2}$ ($P_c^+$) from those with $I_z=-\frac{1}{2}$ ($P_c^0$).
The cross sections decrease dramatically with the increasing transverse momentum. 
 Our study can also tell where to search for the 
 missing hidden charm pentaquarks. The confirmation
 of the complete hidden charm pentaquarks in the heavy quark symmetry
 would further verify their $\Sigma_c^{(*)}\bar{D}^{(*)}$ molecular interpretation.  
 In addition, the relative strength among these cross sections for pentaquarks
 can help us to identify the quantum numbers of the $P_c(4440)$ and $P_c(4457)$.
 \end{abstract}

\maketitle

\section{Introduction}
The successful prediction of the $\Omega$ baryon
has set a milestone of the conventional quark model,
which hints the existence of the color degree of freedom and leads to 
the fundamental theory of the strong interaction, i.e. quantum chromodynamics (QCD). 
 It is also a typical example of the connection between spectroscopy and
 underlying dynamics. 
 The multiquark was first quantitatively studied by Jaffe in 1976~\cite{Jaffe:1976ih} 
 at the budding period of quark model. In the following tens of years, both theorists and experimentalists
 were committed to searching for the missing states and those beyond the conventional quark model, namely exotic states.
The enthusiasm of studying multiquark states, however, was limited by the negative results and the low statistics of the experimental data. 
The situation broke until 2003 by the observation of the $D_{s0}^*(2317)$~\cite{Aubert:2003fg}
  and the $X(3872)$~\cite{Choi:2003ue}. As both the masses of the $D_{s0}^*(2317)$ and the $X(3872)$ are significantly lower than its quark model 
expectation,
they seriously challenged the conventional quark model and served as strong candidates for the exotic states. Up to now, tens of exotic 
  candidates~\cite{Olsen:2017bmm,Yuan:2019zfo} have been observed
   and various proposals were put forward about their configurations from theoretical side~\cite{Chen:2016qju,Chen:2016spr,Esposito:2016noz,Hosaka:2016pey,Dong:2017gaw,Guo:2017jvc,Cerri:2018ypt,Kou:2018nap,Brambilla:2019esw,Guo:2019twa,Liu:2019zoy}.
  
  In a more general concept, all the boson (fermion) hadrons are classified to meson (baryon).
 The statistics of the experimental data for baryon sector is usually
  lower than that for the meson sector due to the number of quarks. Accordingly, the experimental data for exotic baryons are more scarce.
  Research enthusiasm is rekindled by the observation of the first hidden charm pentaquarks
  $P_c(4380)$ and $P_c(4450)$ in the $J/\psi p$ invariant mass distribution of the 
  $\Lambda_b\to J/\psi p K^-$ process~\cite{Aaij:2015tga}.
 As they strongly decay into $J/\psi p$,
they contain as least $c\bar{c}uud$ five quarks unambiguously.
Even before their observation, the hidden charm pentaquarks 
were proposed analogous to the excited baryons in light sector, 
e.g. in Refs.~\cite{Wu:2010jy,Wang:2011rga,Wu:2012md,Xiao:2013yca,Karliner:2015ina,Yang:2011wz,Yuan:2012wz}. 
  An updated analysis~\cite{Aaij:2019vzc} of the LHCb Collaboration with an order-of-magnitude larger data set combined Run-I and II
  shows that the $P_c(4450)$ splits into two narrow peaks $P_c(4440)$
  and $P_c(4457)$, and a third pentaquark $P_c(4312)$ emerges. Various interpretations follow
  this analysis, such as hadronic molecules~\cite{Chen:2019bip,Chen:2019asm,Guo:2019fdo,Liu:2019tjn,He:2019ify,Guo:2019kdc,Shimizu:2019ptd,Xiao:2019mst,Xiao:2019aya,Wang:2019nwt,Meng:2019ilv,Wu:2019adv,Xiao:2019gjd,Voloshin:2019aut,Sakai:2019qph,Wang:2019hyc,Yamaguchi:2019seo,Liu:2019zvb,Lin:2019qiv,Wang:2019ato,Gutsche:2019mkg,Burns:2019iih,Du:2019pij,Wang:2019spc,Xu:2020gjl,Kuang:2020bnk,Peng:2020xrf,Peng:2020gwk,Xiao:2020frg,Dong:2021juy,Peng:2021hkr}, compact pentaquarks~\cite{Ali:2019npk,Zhu:2019iwm,Wang:2019got,Giron:2019bcs,Cheng:2019obk,Stancu:2019qga,Kuang:2020bnk,Ozdem:2021ugy}, 
  hadro-charmonia~\cite{Eides:2015dtr,Eides:2019tgv,Anwar:2018bpu}, and cusp effects~\cite{Kuang:2020bnk}. 
  Among these interpretations,
  the $\Sigma_c^{(*)}\bar{D}^{(*)}$ hadronic molecular picture attracts the most attention based on the fact that
  the $P_c(4312)$ and $P_c(4440)/P_c(4457)$ are close to the $\Sigma_c\bar{D}$ and 
  the $\Sigma_c\bar{D}^*$ thresholds, respectively. According to heavy quark spin symmetry (HQSS), there should exist
  seven hidden charm pentaquarks~\cite{Xiao:2013yca,Du:2021fmf,Xiao:2020frg,Du:2019pij,Pan:2019skd,Liu:2019zvb,Valderrama:2019chc,Liu:2019tjn}, i.e. three $J^P=\frac{1}{2}^-$ pentaquarks, three 
  $J^P=\frac{3}{2}^-$ pentaquarks and one $J^P=\frac{5}{2}^-$ pentaquark with total
  angular momentum $J$ and parity $P$. Three of them 
  are identified as the observed $P_c(4312)$ [$\Sigma_c\bar{D}$], $P_c(4440)$ [$\Sigma_c\bar{D}^*$] and
   $P_c(4457)$[$\Sigma_c\bar{D}^*$] ~\cite{Pan:2019skd,Liu:2019zvb,Valderrama:2019chc,Liu:2019tjn,Du:2019pij,Xiao:2020frg,Du:2021fmf} 
   by the LHCb Collaboration in 2019. 
By fitting to the $J/\psi p$ invariant mass spectrum, it is demonstrated in Refs.~\cite{Du:2019pij,Du:2021fmf} 
that the $\Sigma_c^*\bar{D}$ molecule should correspond to a new narrow $P_c(4380)$
 which leaves a hint for its existence on the $J/\psi p$ distributions. 
 However, the reason for almost invisibility of the $\Sigma_c^*\bar{D}^*$ molecules
  remains to be understood. It could be caused by the small production for the 
  $\Sigma_c^*\bar{D}^*$ channel in the $\Lambda_b^0$ decay compared to other channels. 
  Therefore, searching for the missing pentaquarks in different processes
   is a demanding task to complete the full spectrum and shed light on the underlying dynamics.
 On the other side, their closeness to the $\Sigma_c^{(*)}\bar{D}^{(*)}$ also indicates a large isospin violation
 in their decay rate, for instance the $P_c(4457)\to J/\psi \Delta^+$ process~\cite{Guo:2019fdo}, 
 and a significant deviation of the cross sections of the $P_c^+$ pentaquarks with
 the third isospin component $I_z=+\frac{1}{2}$ from those of their isospin partners $P_c^0$ with $I_z=-\frac{1}{2}$.
  
  Besides the $\Sigma_c^{(*)}\bar{D}^{(*)}$ hadronic molecular picture, the peaking structures in $J/\psi p$ distributions may 
 be caused by kinematical effects, e.g. triangle singularities or cusps~\cite{Aaij:2019vzc,Kuang:2020bnk,Nakamura:2021qvy}. 
 For instance, the $P_c(4457)$ might be generated by the $\Lambda_c^+(2590)\bar{D}^0D_s^{**}$ triangle diagram~\cite{Aaij:2019vzc}. 
The triangle singularities arise when all three hadrons in the triangle-diagrams are nearly on mass shells and are 
manifested as peaks in the mass distributions \cite{Guo:2019twa}. Their manifestation are particularly sensitive to 
the momenta of incoming and outgoing particles. 
 One of the conditions for the triangle singularities is that the mass of the decaying particle, i.e. $\Lambda_b^0$, should be very close to
 the threshold of the connected two internal particles. Once the deviation is larger,
 the triangle singularity condition will not be satisfied and
 the corresponding peaks will disappear. Thus searching for 
 the hidden pentaquark states in the prompt production 
 with a large incoming energy region will help to
 exclude the potential triangle singularity interpretation.
  
  Motivated by the above arguments, we use Madgraph5~\cite{Alwall:2011uj}
  and Phythia8~\cite{Sjostrand:2007gs} to 
  stimulate the prompt production rate of 
  hidden charm pentaquarks in the hadronic molecular picture.
  Our framework is presented in Sec.~\ref{sec:framework}. 
  Results and discussions follows. A brief summary and outlook are given in the last section.
  
\section{Framework}\label{sec:framework}
The inclusive production of a loosely bound $S$-wave hadronic molecule
in hadron collision can be separated into a long-range part and a short-range
 part~\cite{Artoisenet:2009wk,Artoisenet:2010uu,Guo:2014ppa,Guo:2013ufa,Guo:2014sca}
 ~\footnote{The calculation in this work is based on the $\Sigma_c^{(*)}\bar{D}^{(*)}$ molecular picture.
The analogous calculation can also be done for the compact pentaquarks similar to
 the production of the $X(3872)$ in $p\bar{p}$ collision~\cite{Bignamini:2009sk}.},
which is based on the universal scattering amplitude for the low energy scattering.
This kind of separation allows for estimating the cross section of the inclusive production
for a given hadronic molecule, for instance the production of the $X(3872)$~\cite{Artoisenet:2009wk,Artoisenet:2010uu} 
and its bottom analogs~\cite{Guo:2014sca}, the $D_{s0}(2317)$~\cite{Guo:2014ppa},
 the charged $Z_c^{(\prime)\pm}$ and $Z_b^{(\prime)\pm}$\cite{Guo:2013ufa}.
 We employ the formula presented in Refs.~\cite{Guo:2013ufa,Guo:2014sca,Guo:2014sca}
  to estimate the cross sections of inclusive prompt productions of hidden charm pentaquarks
  observed by LHCb in 2019.
 
\subsection{Factorization}
\begin{figure}[htbp]
\begin{center}
 \includegraphics[width=0.40\textwidth]{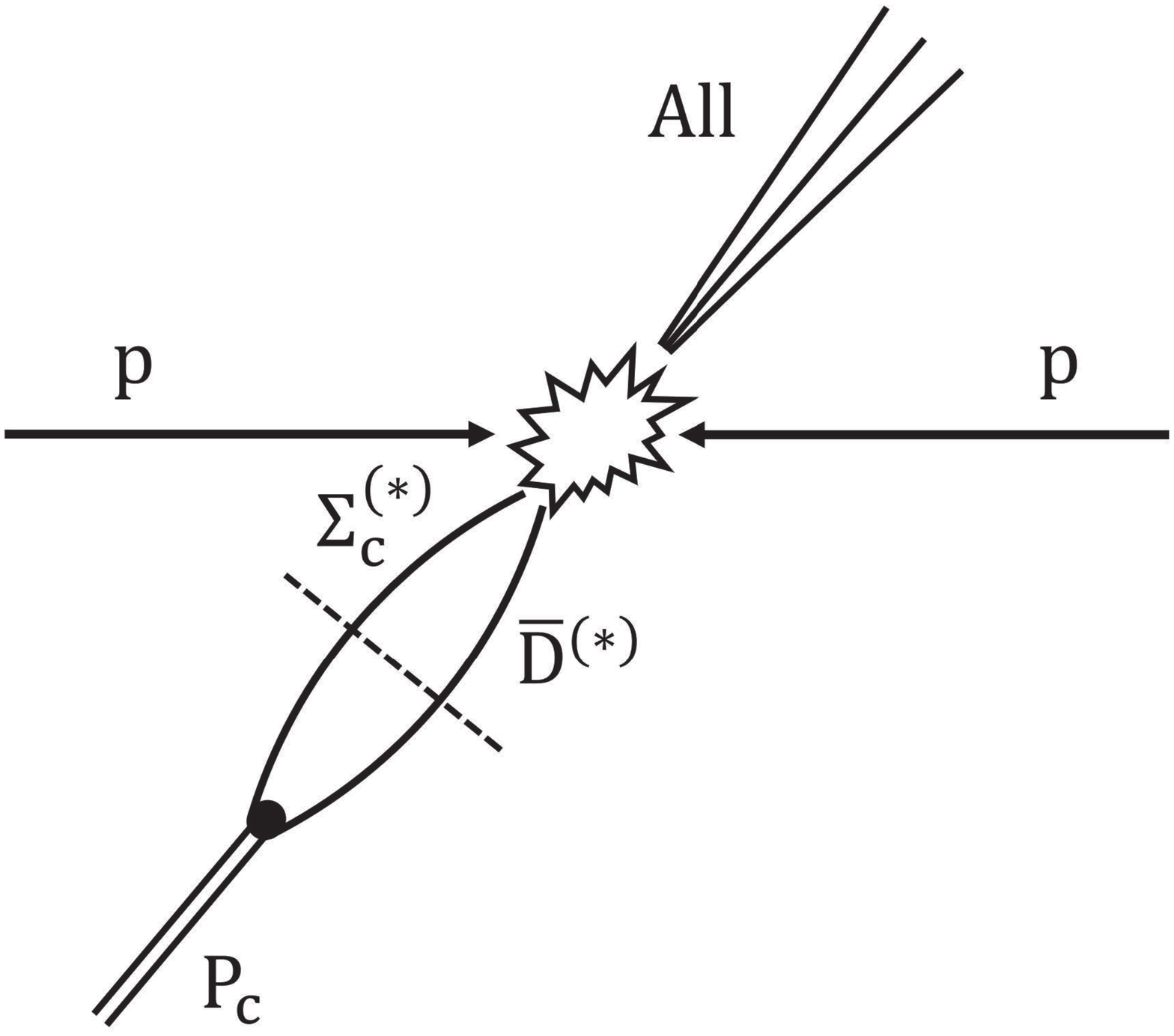}
\caption{The factorization of the inclusive production of hidden charm pentaquarks in $pp$ collision in the
$\Sigma_c^{(*)}\bar{D}^{(*)}$ hadronic molecular picture.  The other particles produced with the $\Sigma_c^{(*)}\bar{D}^{(*)}$ are denoted as ``All" in the figure.}
\label{fig:ppcollision}
\end{center}
\end{figure}

The production amplitude for the inclusive production of hidden charm pentaquarks, in the hadronic molecular picture, 
with small binding energy can be factorized as~\cite{Guo:2013ufa,Guo:2014sca,Guo:2014sca}
\bea
\mathcal{M}[P_{c}(E)]\approx \sum_{\alpha}\int\frac{\text{d}^3\textbf{q}}{(2\pi)^3}\mathcal{M}[\left(\Sigma_{c}^{(*)}\bar{D}^{(*)}\right)^{\alpha}(\textbf{q})+\text{all}]\times G_{\alpha}(E,\textbf{q})\times T_{P_{c}}^{\alpha}(E),
\label{eq:LSE}
\eea
as illustrated by Fig.~\ref{fig:ppcollision}, with $\alpha$ the channel index. 
This factorization is based on the separation of the involved scales, 
i.e. the binding momentum $\sqrt{2\mu E_B}$ with $\mu$ the reduced mass of $\Sigma_{c}^{(*)}\bar{D}^{(*)}$ 
and $E_B$ the binding energy from the large scale of QCD for point-like production in the
effective field theory point of view. While the inclusive production of the 
$\Sigma_{c}^{(*)}\bar{D}^{(*)}$ from $pp$ collision and
the formation of the loosely bound molecular hidden charm pentaquarks
 are regarded as a short-range interaction and 
a long-range interaction, respectively. This idea is proposed in Refs.~\cite{Artoisenet:2009wk,Artoisenet:2010uu}
and has been used to estimate the production of heavy quarkonium-like states
 in Refs.~\cite{Guo:2013ufa,Guo:2014sca,Jin:2016vjn,Albaladejo:2017blx,Wang:2017gay,Braaten:2018eov}
and charm-strange molecules in Ref.~\cite{Guo:2014ppa}
\footnote{The debate of the yield of the $X(3872)$ at high $p_T$ can be found in Refs.~\cite{Guerrieri:2014gfa,Albaladejo:2017blx,Bignamini:2009sk}}.
 In this work, since we only aim at an estimate of 
order-of-magnitude, only the production through $S$-wave $\Sigma_{c}^{(*)}\bar{D}^{(*)}$ channels are considered. 
Here $\mathcal{M}[\left(\Sigma_{c}^{(*)}\bar{D}^{(*)}\right)^{\alpha}+\text{all}]$ is the production
amplitude of the corresponding constituent for a given pentaquark. $T^\alpha$ is the amplitude 
for the $\Sigma_{c}^{(*)}\bar{D}^{(*)}$ to the $P_c$ pentaquark. 
\bea
G_{\alpha}(E,\textbf{q})=\frac{2\mu_{\alpha}}{{q}^{2}-p_{\alpha}^{2}-i\epsilon},
\label{eq:two-body}
\eea
is the intermediated two-body propagator, with $p_\alpha$ the non-relativistic three momentum of the $\alpha$th channel,
extracting from below equation
\bea
p_{\alpha}^{2}\equiv2 \mu_{\alpha}(E-m_{\text{th}}^{\alpha}).
\eea
Here $\mu_\alpha$, $m^\alpha_\text{th}$ and $E$ are the reduced mass, threshold of the $\alpha$th channel and the total energy, respectively. 
In low energy, we are interested in, to the leading order,
the amplitudes $\mathcal{M}[\left(\Sigma_{c}^{(*)}\bar{D}^{(*)}\right)^{\alpha}+\text{all}]$ and $T^\alpha$
could be treated as constants~\cite{Guo:2013ufa,Guo:2014sca,Guo:2014sca}, leaving the equation reduced into 
an algebraic equation. As the result, we only need to calculate the integration  
\bea
\int\frac{\mathrm{d}^{3}\vec{q}}{(2\pi)^{3}} G_{\alpha}(E,\textbf{q}) = \int\frac{\mathrm{d}^{3}\vec{q}}{(2\pi)^{3}} \frac{2\mu_{\alpha}}{\textbf{q}^{2}-p_{\alpha}^{2}-i\epsilon}=\int\frac{q^{2}\mathrm{d}q}{2\pi^{2}}\frac{2\mu_{\alpha}}{q^{2}-p_{\alpha}^{2}-i\epsilon},
\eea
which is linearly divergent and needs to be regularized. To that end, a hard cut-off $\Lambda$ is introduced to render the 
integral well defined
\bea
\mathcal{G}_\alpha(E)\coloneqq\int_{0}^{\Lambda}\frac{q^{2}\mathrm{d}q}{2\pi^{2}}\frac{2\mu_{\alpha}}{q^{2}-p_{\alpha}^{2}-i\epsilon}
=\frac{\mu_\alpha\Lambda}{\pi^2}-\frac{\mu_\alpha p_\alpha}{\pi^2}\mathrm{ArcTanh(\frac{p_\alpha}{\Lambda})}+i\frac{\mu_\alpha p_\alpha}{\pi}.
\label{eq:G}
\eea
%
The value of $\Lambda$ is determined by the effectiveness of the low energy theorem which inherits the non-perturbative mechanism of strong interaction and is of order of 1 $\gev$. 
In our case, we take values $\Lambda=[0.7,1.3]~\gev$
to estimate the cross sections
~\footnote{The lower limit $0.7~\gev$ should be larger than the largest three momentum of the involved dynamic channels. 
The upper limit $1.3~\gev$ is largest value of $\Lambda$ making the physical observables renormalization group 
invariant~\cite{Du:2021fmf}.}. 

Before going into the production of the $P_c$ states,
the cross section of the inclusive $\Sigma_{c}^{(*)}\bar{D}^{(*)}$ production
should be estimated by Monte Carlo (MC) simulation and reads as
\bea
\mathrm{d}\sigma[\Sigma_{c}^{(*)}\bar{D}^{(*)}]_{MC}=K_{\Sigma_{c}^{(*)}\bar{D}^{(*)}}\frac{1}{\text{flux}}\sum_{\text{all}}\mathrm{d}\phi_{\Sigma_{c}^{(*)}\bar{D}^{(*)}+\text{all}}|\mathcal{M}[\Sigma_{c}^{(*)}\bar{D}^{(*)}+\text{all}]|^{2}\frac{\mathrm{d}^{3}k}{(2\pi)^{3}2\mu},
\eea
with $\mathrm{d}\phi_{\Sigma_{c}^{(*)}\bar{D}^{(*)}+\text{all}}$ the phase space of the $\Sigma_{c}^{(*)}\bar{D}^{(*)}$
and all the other particles. $k$ is the relative momentum between $\Sigma_{c}^{(*)}\bar{D}^{(*)}$ in its center-of-mass frame. 
$K_{\Sigma_{c}^{(*)}\bar{D}^{(*)}}$ is a normalization factor 
to compensate the difference between the MC and experimental data,
and is taken the value $1$ as an order-of-magnitude estimate.  
In total, the cross section of the inclusive $P_{c}$ production can be written as
\bea\label{eq:prodPc}
\sigma[P_{c}]=\sum_\alpha\frac{1}{4m_{\Sigma_{c}^{(*)}}m_{\bar{D}^{(*)}}}|\tilde{g}^{P_c}_{\alpha}|^{2}|\mathcal{G}_{\alpha}|^{2}\left(\frac{\mathrm{d}\sigma[\left(\Sigma_{c}^{(*)}\bar{D}^{(*)}(k)\right)^{\alpha}]}{\mathrm{d}k}\right)_{MC}\frac{4\pi^{2}\mu_\alpha}{k^{2}},
\eea
where
\bea
|\tilde{g}^{P_c}_\alpha|^2=\frac{(2J_{P_c}+1)|g_\alpha^{P_c}|^2}{\sum_i(2J_i+1)|g_\alpha^i|^2}
\label{eq:efcoupling}
\eea
means the fraction
of the $\alpha$th channel events to a given $P_c$ with spin $J_{P_c}$. 
The sum $i$ in the denominator runs all the pentaquark states which 
couple to the $\alpha$th channel. 
$g_\alpha$ is the coupling of the $\alpha$th channel
to a given $P_c$ state and the values can be found in App.~\ref{app:couplings}.

\subsection{Monte Carlo Simulation}
The production of hidden charm pentaquarks in the hadronic 
molecular picture should follow the production of the corresponding
constituents, i.e. the heavy quark pair $c\bar{c}$ in the parton level. 
Considering the other produced particles in the inclusive process, a third parton should be
produced simultaneously.  As the result, the $2\to 3$ parton process should be 
generated through hard scattering and hadronized into final hadrons via non-perturbative mechanism.

\begin{figure}[htbp]
\begin{center}
 \includegraphics[width=0.45\textwidth]{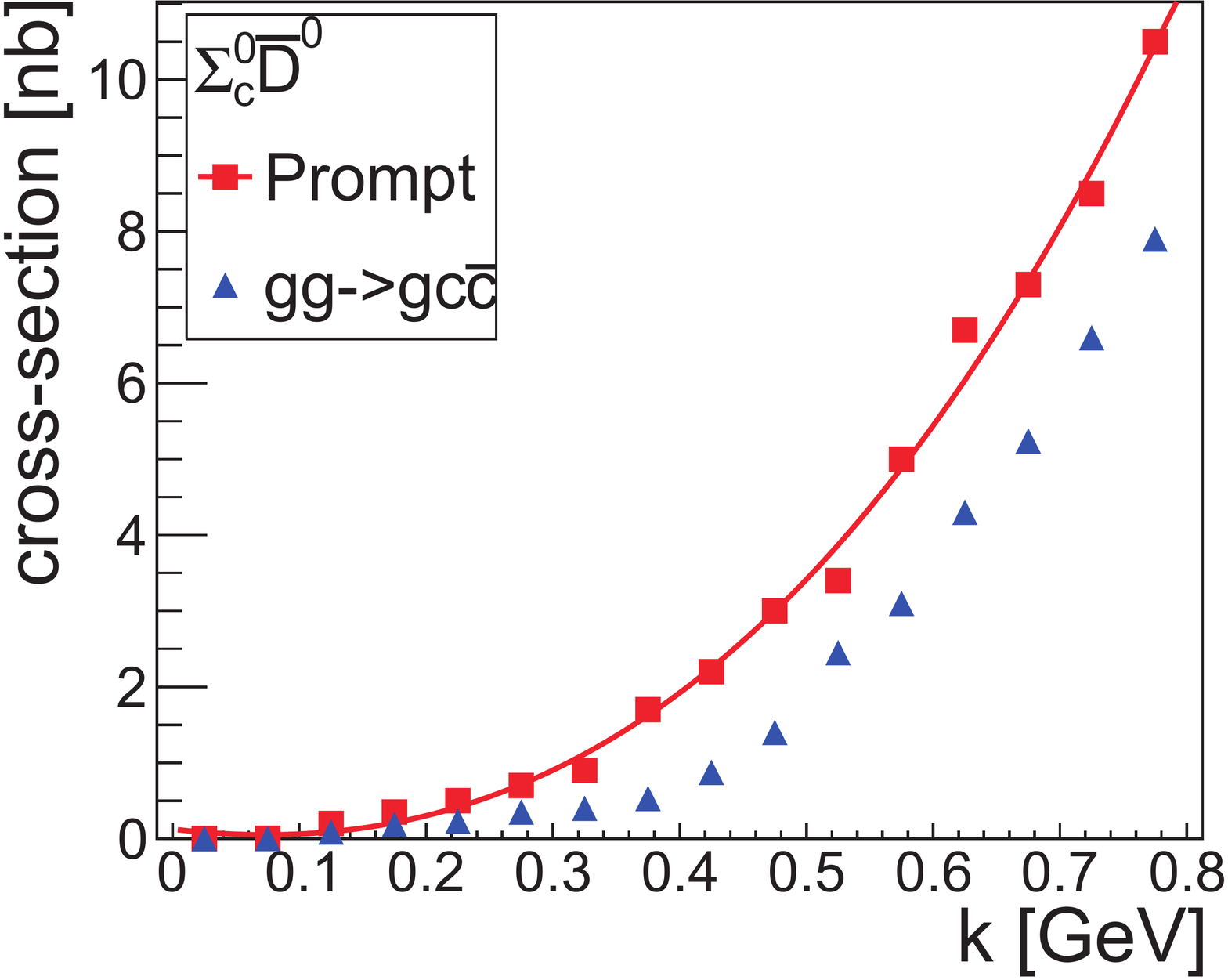} 
\caption{The cross section of the inclusive $\Sigma_c^0\bar{D}^0$ pair production in 
terms of their relative momentum at the center-of-mass energy $\sqrt{s}=7~\tev$.
The red box points are obtained by all the parton level diagrams switched on.
The blue triangle points are those for only the $gg\to gc\bar{c}$ switched on. 
The red sold curve is an interpolation.
The case for other $\Sigma_c^{(*)}\bar{D}^{(*)}$ channel is the same, 
i.e. the $gg\to gc\bar{c}$ parton process is the most important contribution. 
} 
\label{fig:ggtogcc}
\end{center}
\end{figure}

Similar to those in Refs.~\cite{Guo:2013ufa,Guo:2014sca,Guo:2014sca}, we generate the $2\to 3$ process via
Madgraph5~\cite{Alwall:2011uj} and use Phythia8~\cite{Sjostrand:2007gs} for the hadronization. 
As the two constituents $\Sigma_c^{(*)}\bar{D}^{(*)}$ should be collinear and with relative small momentum,
the cut $p_\text{T}>3.5~\gev$ and $|y|<2$ are implemented for the heavy quark pair. In principle, all the parton level $2\to 3$
processes will contribute. However, we demonstrate numerically, as shown in Fig.~\ref{fig:ggtogcc}, that 
the $gg\to gc\bar{c}$
\footnote{To estimate the uncertainty of this approximation, we compare the $\Sigma_c^+\bar{D}^0$
inclusive cross sections within these two frameworks and find that  the deviation is under $5\%$. }
 is the most important parton process similar to its contribution in the $X(3872)$ production~\cite{Artoisenet:2009wk,Artoisenet:2010uu}.
Accordingly, to improve the efficiency of the code, 
only the $gg\to gc\bar{c}$ process is considered in Madgraph5.
In addition, the dependence on the relative momentum $k$
is~\cite{Guo:2013ufa,Guo:2014sca,Guo:2014sca}
\bea
\frac{\mathrm{d}\sigma[\Sigma_{c}^{(*)}\bar{D}^{(*)}(k)]_{\text{MC}}}{\mathrm{d}k}\sim k^{2}
\label{eq:kdep}
\eea
at low energy without considering the final-state-interaction (FSI).
\begin{figure}[htbp]
\begin{center}
\includegraphics[width=0.5\textwidth]{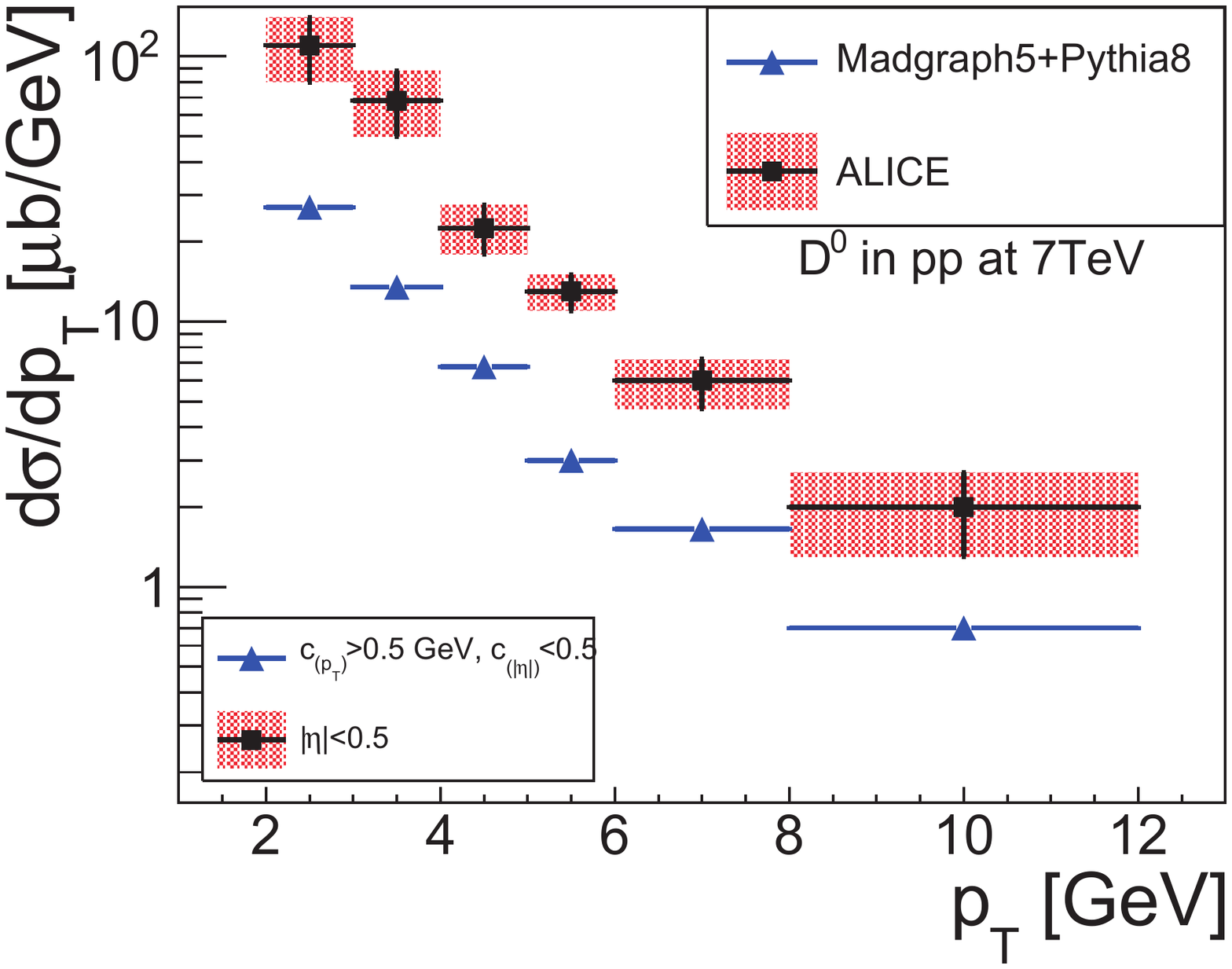} \hspace{-1.5cm}\includegraphics[width=0.5\textwidth]{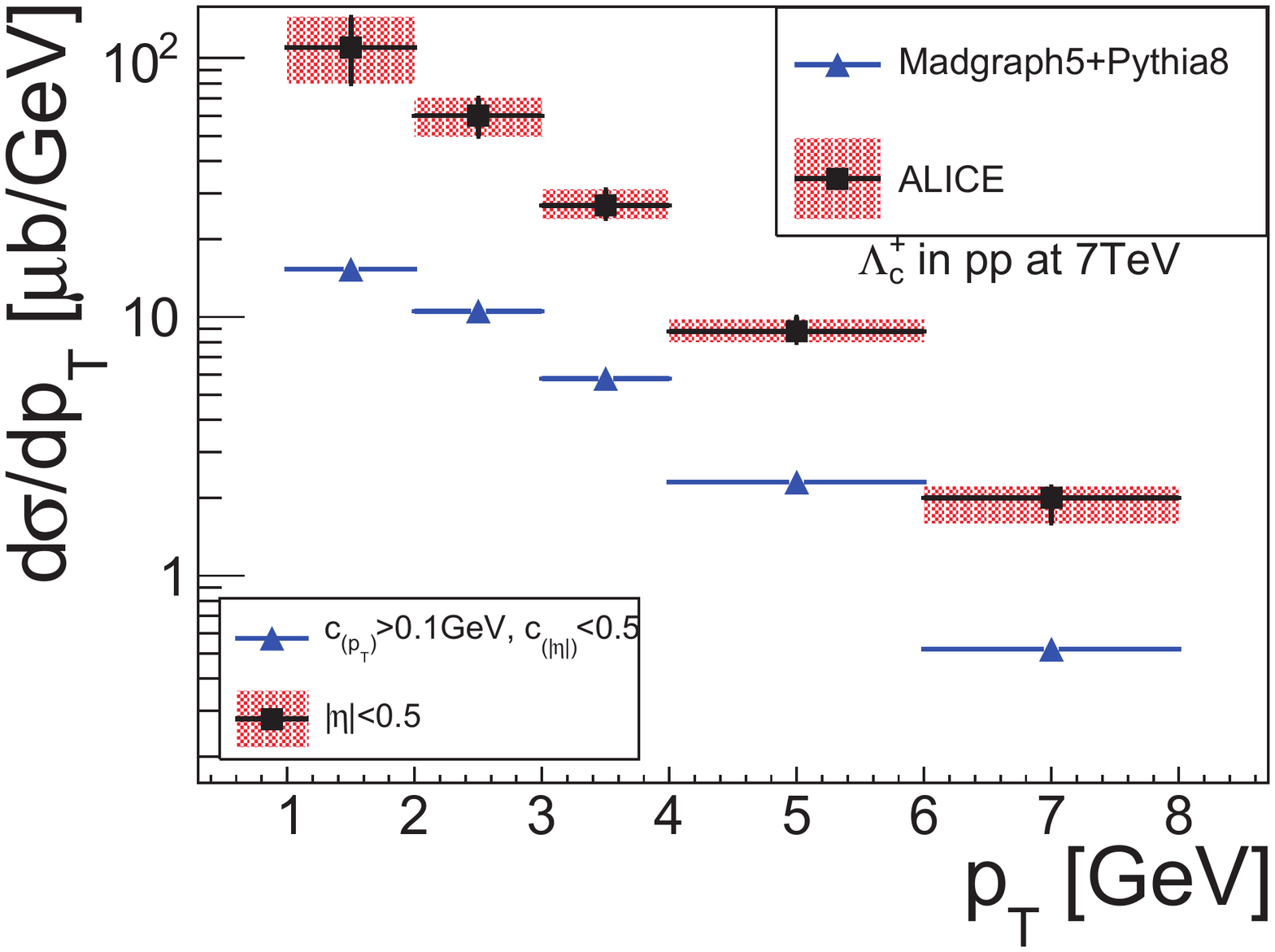}
\caption{MC simulation of the cross section of the inclusive $D^0$ (left) and $\Lambda_c^+$ (right) production 
in the $pp$ collision at $\sqrt{s}=7~\tev$ comparing with
the experimental data from Refs.~\cite{Tieulent:2011hn,Acharya:2017kfy}. 
The rapidity cut $|\eta|<0.5$ is implemented for charm quark
to compare with the experimental data. The black box and blue triangle points
are for the experimental data and MC results, respectively.}
\label{fig:Comparision}
\end{center}
\end{figure}

As the hadronization process is still unclear and model-dependent,
the hadronization implemented in Phythia8~\cite{Sjostrand:2007gs}
is incomplete and we might underestimate the yield of heavy hadron pairs.
A comparison between the MC simulation results, with all the parton processes considered,
 of the $D^0$ meson and the $\Lambda_c^+$ charm baryon\footnote{As the cross section of the $\Sigma_c$ charmed baryon is not existing, we use that of the $\Lambda_c^+$ as an illustration for the production of charm baryons.} 
with the experimental data are presented in Fig.~\ref{fig:Comparision}.
Although the experimental data and the MC simulation result are of the same order,
the deviation is still sizable, especially for the charmed baryon. 
The deviation is because of the missing dynamics in Phythia8,
for instance the feed-down charm meson/baryon from bottom meson/baryon~\cite{Cacciari:2012ny}. 
As we only make an order-of-magnitude estimate for the cross sections, this deviation 
can be accepted.

\section{Results and Discussions}\label{sec:results}
\begin{figure}[htbp]
\begin{center}
\includegraphics[width=0.45\textwidth]{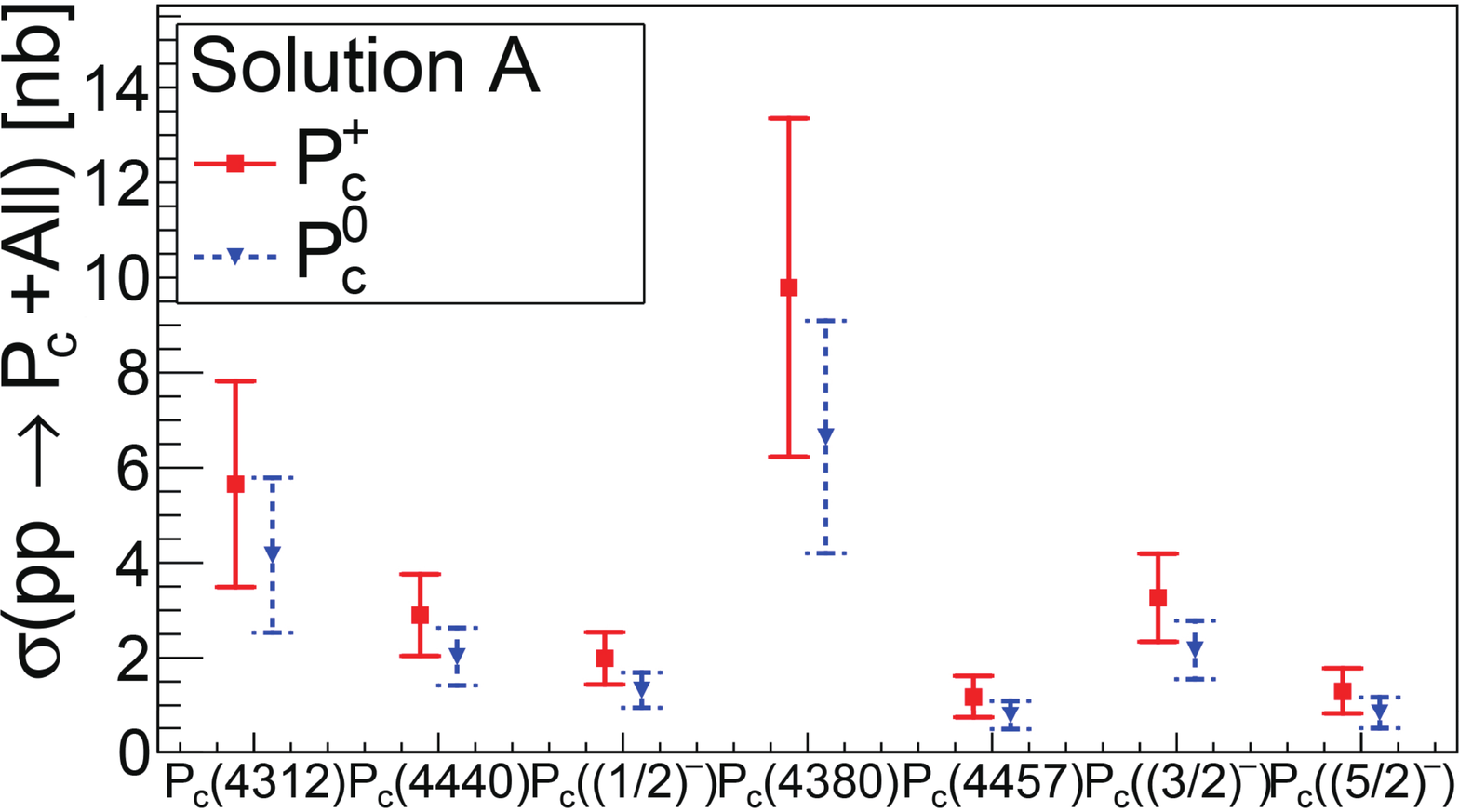}\hspace{0.5cm}\includegraphics[width=0.45\textwidth]{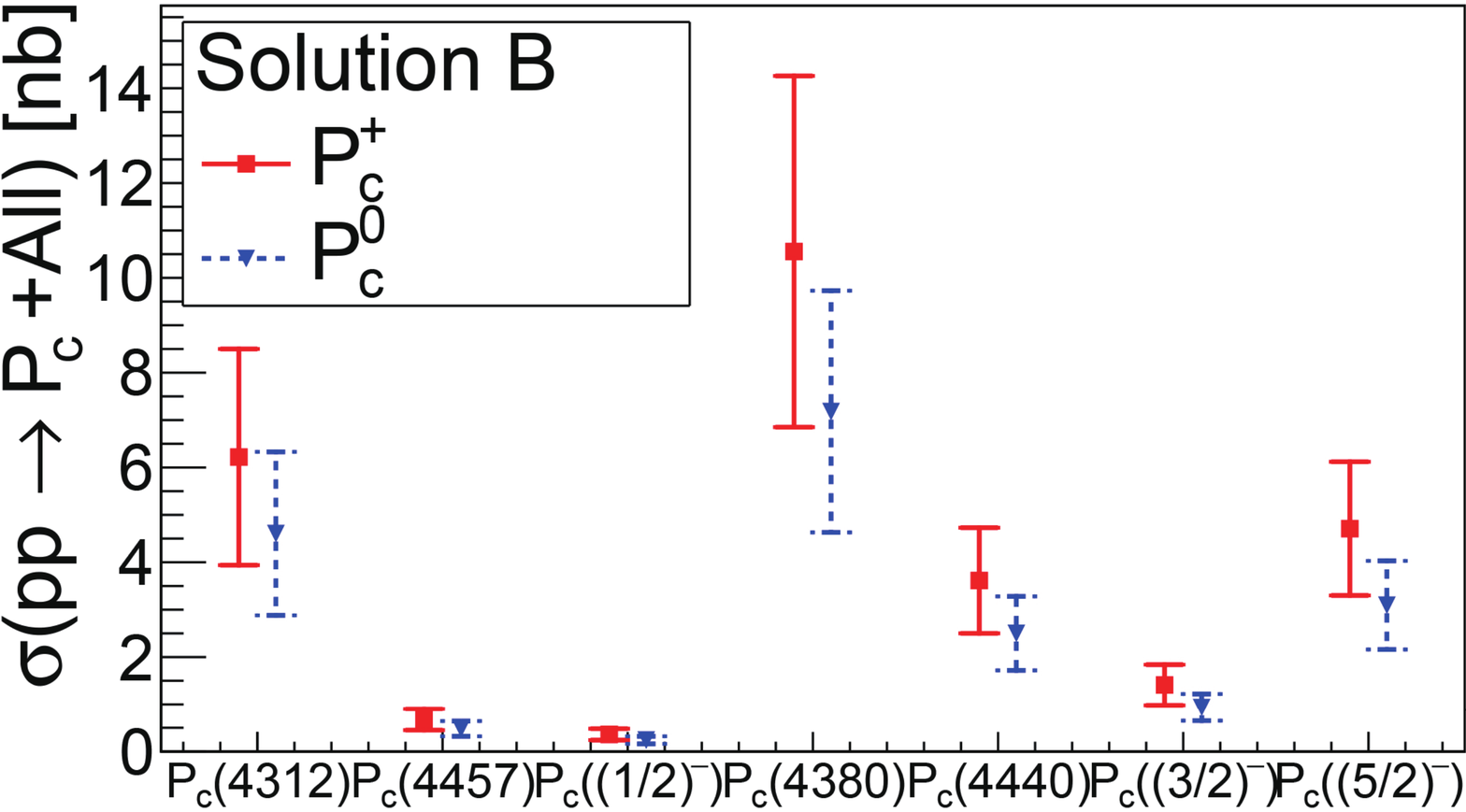}
\caption{The cross section of the prompt production of the hidden charm 
pentaquarks in the $pp$ collision at $\sqrt{s}=7~\tev$.
The red boxes and blue inverse triangles are for the charged pentaquarks $P_c^+$
and the neutral ones $P_c^0$, respectively.
The left and right panels are for the solution A and solution B in the hadronic molecular picture.
The lower limits, the upper ones and the central values correspond
 to the cross sections with $\Lambda=0.7~\gev$, $\Lambda=1.3~\gev$
 and their average values. The difference of uncertainties are mainly
  because of the $k$-dependence of the heavy hadron pairs cross section, 
   e.g. Eq.~\eqref{eq:kdep} in MC simulation. In our calculation, the $k$-dependence in each $p_\mathrm{T}$
   bin is evaluated numerically, but not the simple $k^2$ behavior.
}
\label{fig:crosssection}
\end{center}
\end{figure}

As discussed in the introduction, the closeness of the $P_c$s
to the $\Sigma_c^{(*)}\bar{D}^{(*)}$ threshold might indicate a 
significant deviation of the cross sections of the $P_c^+$
from those of the $P_c^0$, we explicitly consider the cross sections of the inclusive productions of 
$P_c^+$ and $P_c^0$
 with wave functions~\footnote{The third isospin components of $P_c^+$ and $P_c^0$ are $I_z=+\frac{1}{2}$ and $I_z=-\frac{1}{2}$, respectively. In what follows, if the charged property of the hidden charm pentaquarks
is not specified, the argument works for both of them.}
\bea
P_c^+&=&-\sqrt{\frac{1}{3}}\Sigma_c^{(*)+}\bar{D}^{(*)0}+\sqrt{\frac{2}{3}}\Sigma_c^{(*)++}D^{(*)-},\\
P_c^0&=&\sqrt{\frac{1}{3}}\Sigma_c^{(*)+}D^{(*)-}-\sqrt{\frac{2}{3}}\Sigma_c^{(*)0}\bar{D}^{(*)0},
\eea
in the hadronic molecular picture. To study the deviation quantitatively,
the ratio 
\bea
\text{Ratio} (P_c^+/P_c^0)\equiv\frac{\mathrm{d}\sigma (P_c^+)}{\mathrm{d}p_\text{T}}/\frac{\mathrm{d}\sigma (P_c^0)}{\mathrm{d}p_\text{T}}
\label{eq:ratio}
\eea
\begin{figure}[htbp]
\begin{center}
\includegraphics[width=7.7cm,height = 6.8cm]{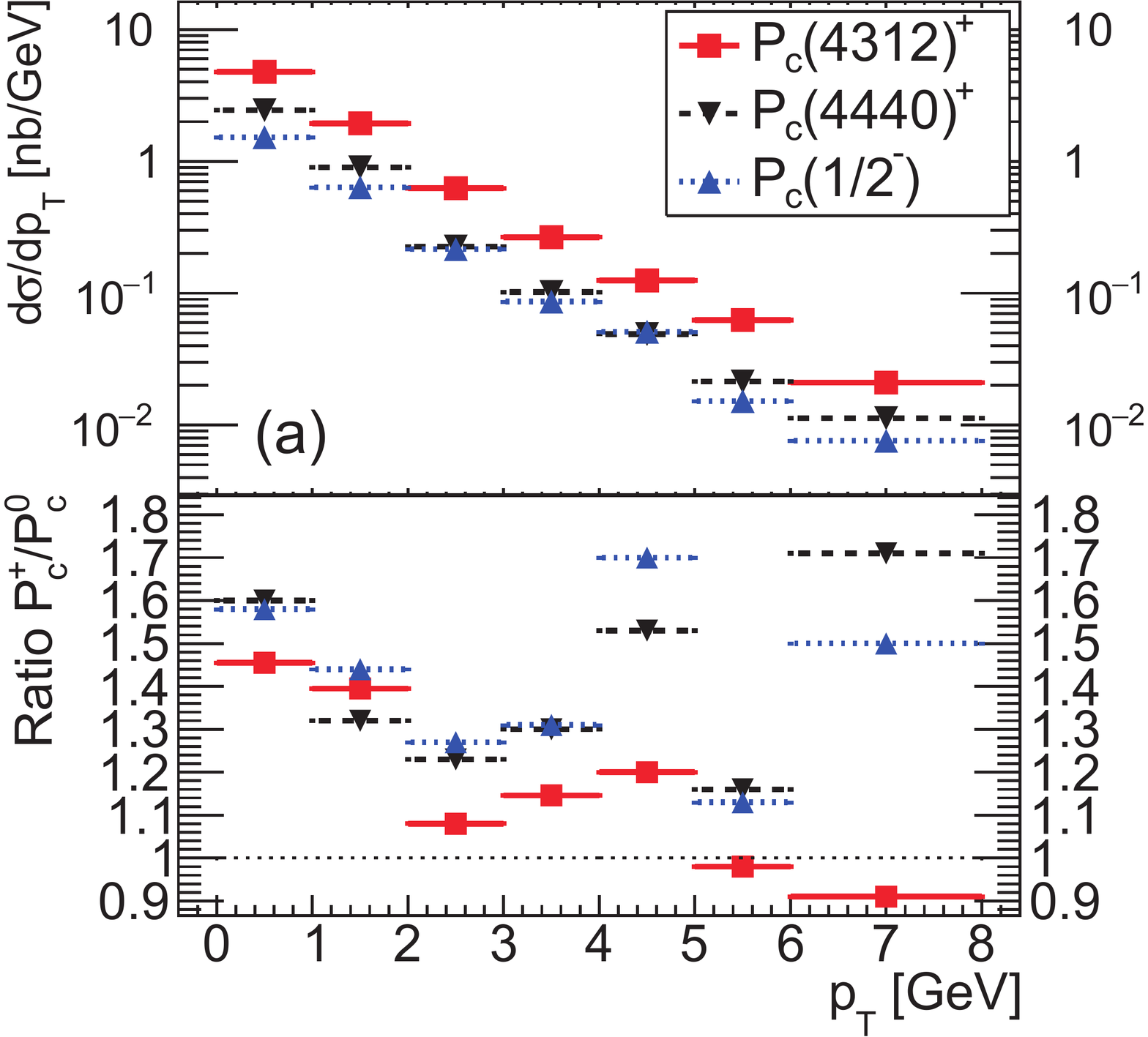}
\includegraphics[width=7.7cm,height = 6.8cm]{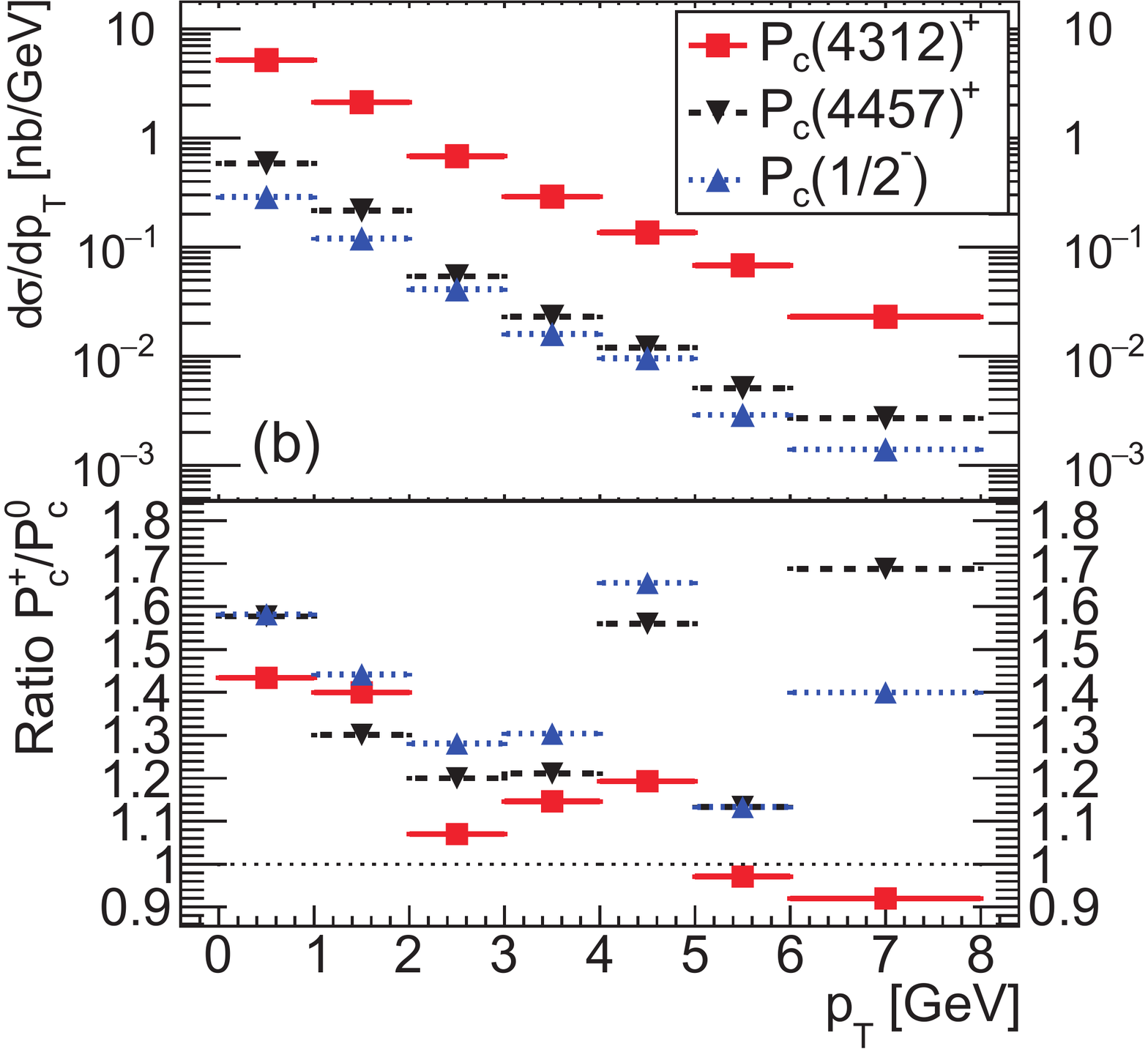}\\
\includegraphics[width=7.7cm,height = 6.8cm]{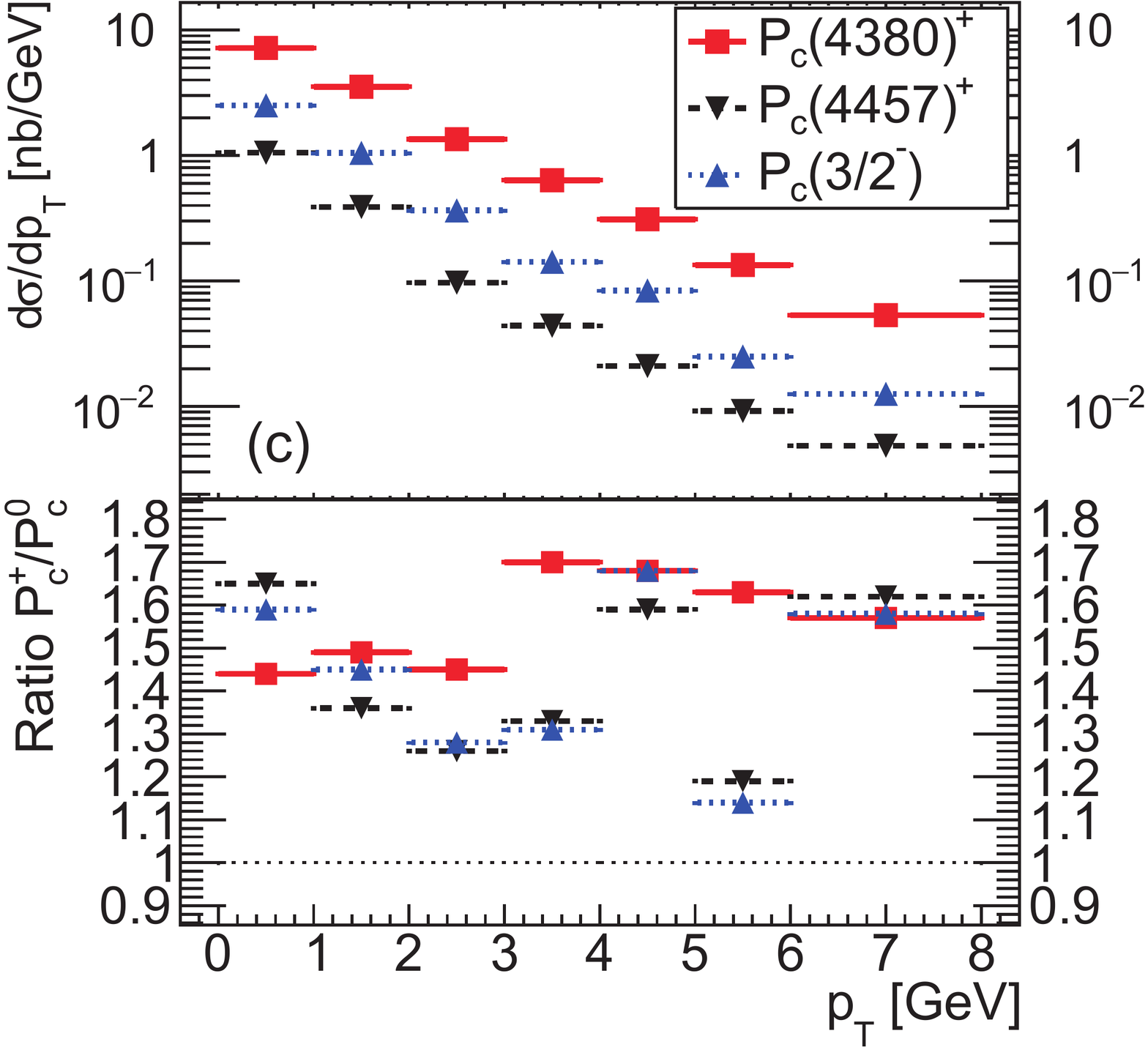}
\includegraphics[width=7.7cm,height = 6.8cm]{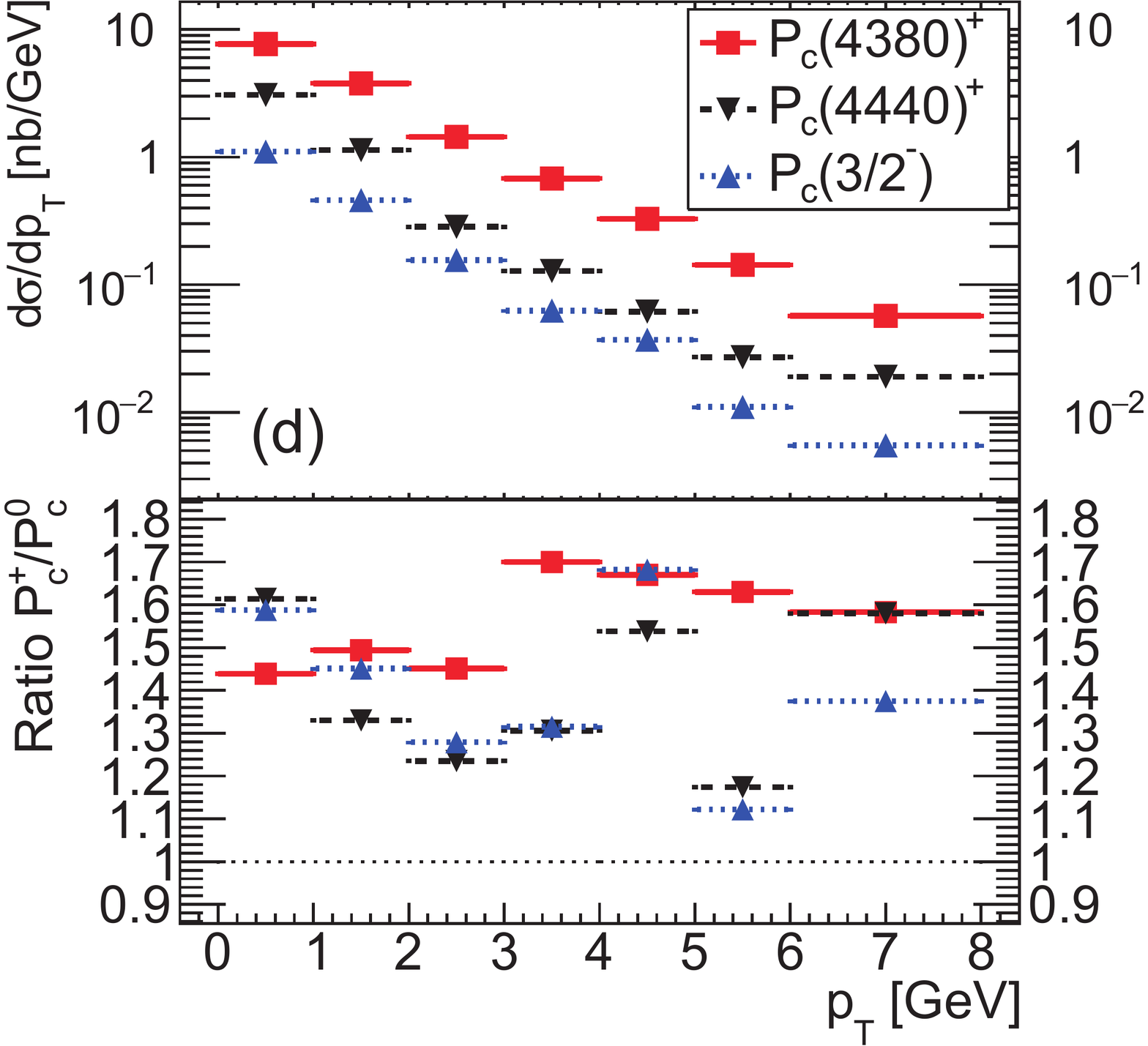}\\
\includegraphics[width=7.7cm,height = 6.8cm]{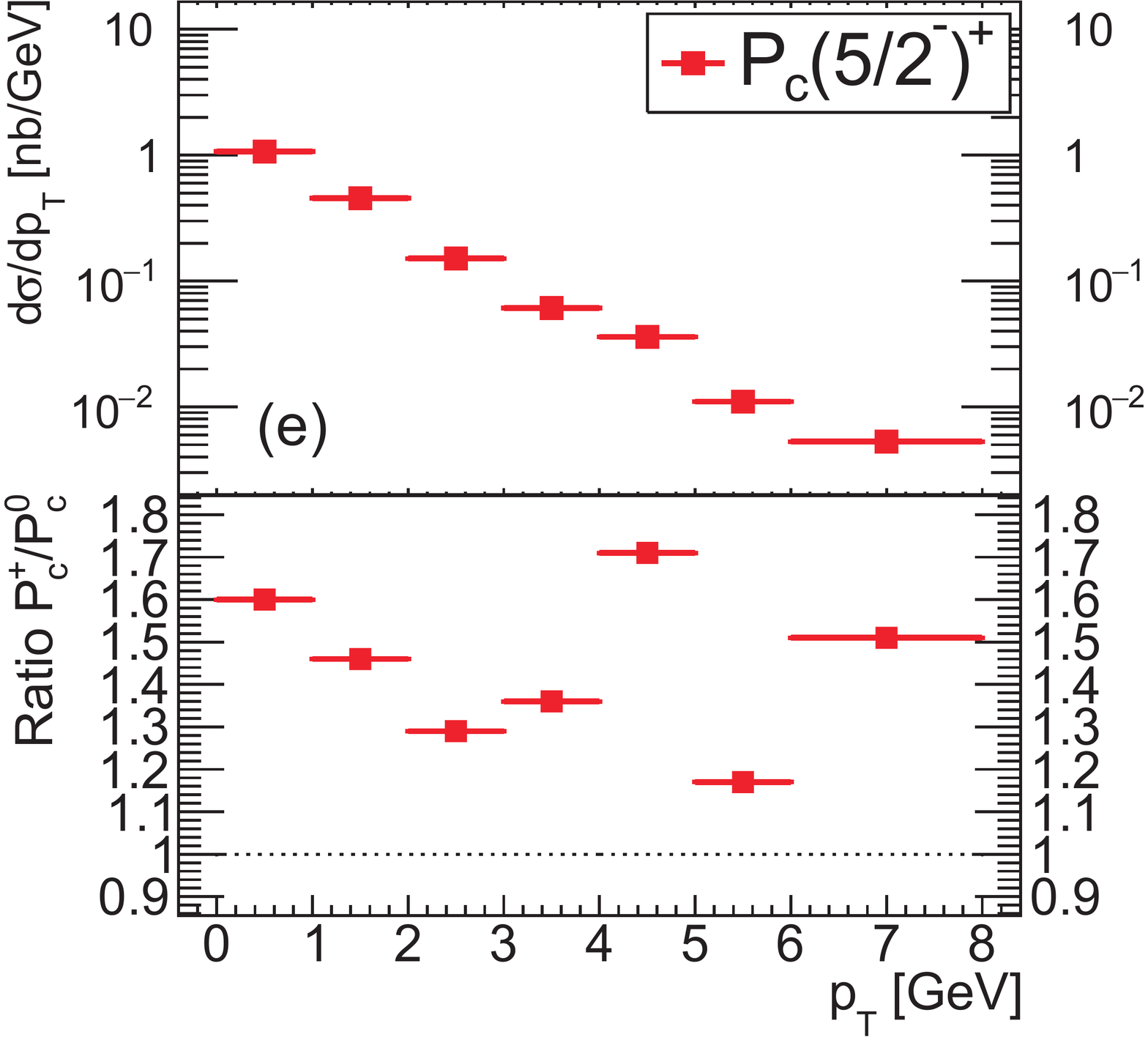}
\includegraphics[width=7.7cm,height = 6.8cm]{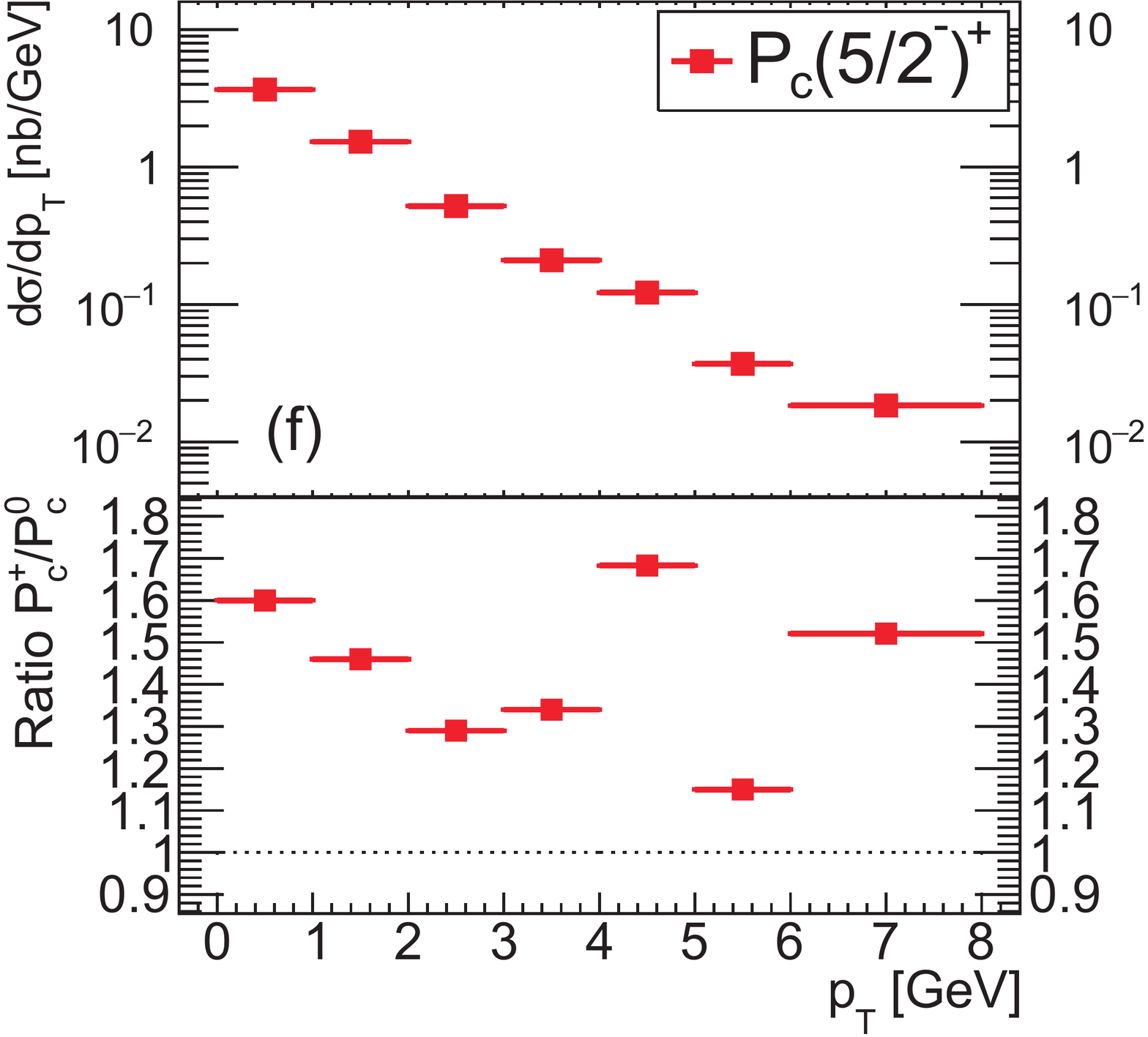}
\caption{The cross sections of the inclusive charged
hidden charm pentaquarks, i.e. the charged $P_c^+$s, in $pp$ collision at $\sqrt{s}=7~\tev$
in terms of the transverse momentum $p_\text{T}$ with $\Lambda=0.7~\gev$.
Their ratios relative to the neutral ones, as defined in Eq.~\eqref{eq:ratio}, are 
also presented below each figure.
The three figures on the left (a,c,e) and right (b,d,f) hand sides are for the solution A and B in Ref.~\cite{Du:2021fmf}, respectively.
Figures (a), (b) are for the $P_c$s with quantum number $J^P=\frac{1}{2}^-$.
Figures (c), (d) are for the $P_c$s with quantum number $J^P=\frac{3}{2}^-$.
Figures (e), (f) are for the $P_c$s with quantum number $J^P=\frac{5}{2}^-$.
In each figure, the $P_c$s are labeled, from lower to higher mass, as red box, black inverted triangle and blue triangle.}
\end{center}
\label{fig:Pc}
\end{figure}
\begin{table}
\caption{The estimate of the inclusive cross section of hidden charm pentaquark states $P_c^+$s
and $P_c^0$s in $pp$ collision at $\sqrt{s}=7~\tev$ with the lower and upper limits
 corresponding to $\Lambda=0.7~\gev$ and $\Lambda=1.3~\gev$, respectively.}
	\begin{tabular}{|c|c|c|c|c|c|c|}
		\hline 
		$J^{P}$ & \multicolumn{3}{c|}{Solution A} & \multicolumn{3}{c|}{Solution B}\\
		\hline 
		\hline 
		\multirow{4}{*}{$\frac{1}{2}^{-}$} & states & $P_{c}^{+}(\mathrm{nb})$ & $P_{c}^{0}(\mathrm{nb})$ & states & $P_{c}^{+}(\mathrm{nb})$ & $P_{c}^{0}(\mathrm{nb})$\\
		\cline{2-7} \cline{3-7} \cline{4-7} \cline{5-7} \cline{6-7} \cline{7-7} 
		& $P_{c}(4312)$ & $3.49\sim 7.82$ & $	2.53 \sim 5.79$ & $P_{c}(4312)$ & $3.94\sim 8.50$ & $2.88\sim 6.33$\\
		\cline{2-7} \cline{3-7} \cline{4-7} \cline{5-7} \cline{6-7} \cline{7-7} 
		& $P_{c}(4440)$ & $2.04 \sim 3.76 $ & $ 1.42 \sim 2.63 $ & $P_{c}(4457)$ & $0.45  \sim 0.90 $ & $ 0.32\sim 0.64$\\
		\cline{2-7} \cline{3-7} \cline{4-7} \cline{5-7} \cline{6-7} \cline{7-7} 
		& $P_{c}(\frac{1}{2}^{-})$ & $1.44\sim 2.54$ & $0.95 \sim 1.69$ & $P_{c}(\frac{1}{2}^{-})$ & $0.24 \sim 0.48 $ & $0.16 \sim 0.32 $\\
		\hline 
		\multirow{3}{*}{$\frac{3}{2}^{-}$} & $P_{c}(4380)$ & $6.23\sim 13.35 $ & $ 4.20 \sim 9.09 $ & $P_{c}(43
		80)$ & $6.85 \sim 14.26$ & $4.63\sim9.73$\\
		\cline{2-7} \cline{3-7} \cline{4-7} \cline{5-7} \cline{6-7} \cline{7-7} 
		& $P_{c}(4457)$ & $0.73\sim 1.62$ & $0.48\sim 1.09 $ & $P_{c}(4440)$ & $2.50\sim4.73$ & $1.72\sim3.28$\\
		\cline{2-7} \cline{3-7} \cline{4-7} \cline{5-7} \cline{6-7} \cline{7-7} 
		& $P_{c}(\frac{3}{2}^{-})$ & $2.34\sim 4.19$ & $1.55 \sim 2.78$ & $P_{c}(\frac{3}{2}^{-})$ & $0.98\sim1.84$ & $0.65\sim1.22$\\
		\hline 
		$\frac{5}{2}^{-}$ & $P_{c}(\frac{5}{2}^{-})$ & $0.81\sim1.78$ & $0.50\sim1.17$ & $P_{c}(\frac{5}{2}^{-})$ & $3.30\sim 6.12$ & $2.16\sim4.03$\\
		\hline 
	\end{tabular}
\label{tab:cs}
\end{table}
between the cross sections of the $P_c^+$ and $P_c^0$ is also defined, where
the statistic uncertainties have been cancelled out.
The cross sections and the corresponding ratios for the inclusive production of the $P_c$s
in $pp$ collision at center-of-mass energy $\sqrt{s}=7~\tev$ are presented in Fig.~\ref{fig:Pc}.
The left and right panels in Fig.~\ref{fig:Pc} are for the two solutions, denoted as A and B, in the molecular picture \cite{Liu:2019tjn,Liu:2019zvb,Du:2019pij,Du:2021fmf}.  In Solution A, the $P_c(4440)$ and the $P_c(4457)$ are assigned as $\frac{1}{2}^-$ and $\frac{3}{2}^-$ pentaquarks, respectively. In Solution B, they are interchanged. From Fig.~\ref{fig:Pc}, one can see that the production rates decrease dramatically
with the increasing $p_\text{T}$, which stems from the decreasing behavior of the fragmentation functions~\cite{Artoisenet:2009wk}. 
 The significant deviations of the cross sections of the charged $P_c^+$s
from those of the neutral ones $P_c^0$s 
 can also be seen directly from cross section of their constituents in App.~\ref{app:SigmacDbar}.
The total cross sections are several $\mathrm{nb}$
as shown in Table~\ref{tab:cs} and Fig.~\ref{fig:crosssection}, with the lower and upper limits corresponding to those with
$\Lambda=0.7~\gev$ and $\Lambda=1.3~\gev$ as an order-of-magnitude estimate. 
Considering the integrated luminosity $\mathcal{L}=34~\mathrm{nb}^{-1}$~\cite{Aad:2011dr} of LHC at $\sqrt{s}=7~\tev$,
we would expect to collect several tens $P_c$s events in LHC. 
Furthermore, the observation of the narrow $P_c(4380)$ \cite{Du:2019pij,Du:2021fmf} and the $P_c$s relevant to the $\Sigma_c^*\bar{D}^*$
channel will complete the spectroscopy of the hidden charm pentaquarks 
of the $\Sigma_c^{(*)}\bar{D}^{(*)}$ hadronic molecular
 picture~\cite{Du:2021fmf,Xiao:2020frg,Du:2019pij,Pan:2019skd,Liu:2019zvb,Valderrama:2019chc,Liu:2019tjn}.
 %

The result of two solutions for the $P_c$ productions are collected in Table~\ref{tab:cs} and Fig.~\ref{fig:crosssection}. In both solutions, the cross section for $P_c(4380)$ is the largest one due to the largest production of $\Sigma_c^*\bar{D}$ channel (Eq.~\eqref{eq:relation_of_cs}) and its strong coupling to $P_c(4380)$, which makes the prompt production in $pp$ collision an ideal platform for the search of the narrow $P_c(4380)$~\cite{Du:2019pij,Du:2021fmf}.  For $J^P=1/2^-$ channel, 
  \bea
    \label{eq:A12}
  \sigma_{\text{A}}\left(P_{c}(4312)\right)&>&\sigma_{\text{A}}\left(P_{c}(4440)\right)>\sigma_{\text{A}}\left(P_{c}(\frac{1}{2}^{-})\right) ,\\
  \sigma_{\text{B}}\left(P_{c}(4312)\right)&>&\sigma_{\text{B}}\left(P_{c}(4457)\right)>\sigma_{\text{B}}\left(P_{c}(\frac{1}{2}^{-})\right),
  \label{eq:B12}
  \eea
  with subindexes A and B for the two solutions.  As they mainly couple to the 
  $\Sigma_c\bar{D}$, $\Sigma_c\bar{D}^*$, $\Sigma_c^*\bar{D}^*$ channels, respectively,
  the relations Eqs.~\eqref{eq:A12} and ~\eqref{eq:B12} are largely determined by the cross sections of their 
  constituents
  \bea
\sigma\left(\Sigma_{c}\bar{D}\right)>\sigma\left(\Sigma_{c}\bar{D}^{*}\right)\backsimeq\sigma\left(\Sigma_{c}^{*}\bar{D}^{*}\right).
\eea
For the $\Sigma_{c}^{*}\bar{D}^{*}$ relevant $P_c$ states, 
a further suppressed factor comes from the denominator of
 Eq.~\eqref{eq:efcoupling}, i.e.  
three $\frac{1}{2}^-$, three $\frac{3}{2}^-$ and one $\frac{5}{2}^-$ states coupling to the same channel.
In addition, one can also see the relation
  \bea
\sigma\left(P_{c}(4440)\right)&>&\sigma\left(P_{c}(4457)\right)
  \eea
in both solutions. That is because the cross section of a
     state is proportional to the absolute square of its coupling to the constitutes and thus its binding energy 
     for shallow bound states~\cite{Guo:2017jvc}. 
     
From Table~\ref{tab:cs} and Fig.~\ref{fig:crosssection}, one can also see different patterns of the prompt cross sections for Solutions A and B which can be used to distinguish the two solutions. It is noticed that the productions of the $P_c$ states are sensitive to the cross sections of the $\Sigma_c^{(*)}\bar{D}^{(*)}$ which however receive large uncertainties due to the unknown hadronisation mechanism. To some extend, the results in Table~\ref{tab:cs} and Fig.~\ref{fig:crosssection} are understood as an order-of-magnitude estimate and a comparison of the cross sections between different $P_c$ states should be made with great care. 
Among them the situation for the $J^P=\frac{5}{2}^-$ pentaquark is the most simple one,
as it only couples to the $\Sigma_c^*\bar{D}^*$ channel. That
the cross section in Solution A is much smaller than that in Solution B is due
to the smaller distance of their pole positions to the $\Sigma_c^*\bar{D}^*$ threshold and the resulting smaller effective coupling.
An important observation is that the three $\Sigma_c^*\bar{D}^*$ molecular states exhibit different patterns in two solutions. For Solution B, the state with higher spin has a larger binding energy, and thus effective coupling. Combined with the enhancing factor $(2J+1)$ in Eq.~\eqref{eq:efcoupling} for higher spins, it leads to a significant relation, see e.g. in Fig.~\ref{fig:crosssection}, 
\bea
\sigma_{\text{B}}\left(P_c(\frac{1}{2}^-)\right)<\sigma_{\text{B}}\left(P_c(\frac{3}{2}^-)\right)<\sigma_{\text{B}}\left(P_c(\frac{5}{2}^-)\right).
\eea
However, the enhancing factor for the higher spin is roughly balanced by the smaller effective coupling in Solution A, which makes the cross sections for the three states comparable. We stress that this difference is independent on the production of $\Sigma_c^{(*)}\bar{D}^{(*)}$ and only relies on the mass of pattern of the three $\Sigma_c^*\bar{D}^*$ states. It provides us an important way to distinguish the two solutions, 
as well as to identify the quantum numbers of the the $P_c(4440)$ and $P_c(4457)$
 since they are assigned to $J^P=\frac{1}{2}^-$ and  $\frac{3}{2}^-$ in Solution A and interchanged in Solution B.
     
\section{Summary}
In a short summary,  within the $\Sigma_c^{(*)}\bar{D}^{(*)}$
hadronic molecular picture, we estimate the prompt cross sections 
of the seven hidden charm pentaquarks at center-of-mass energy $\sqrt{s}=7~\mathrm{TeV}$ in the $pp$ collision.
The observation of the narrow $P_c(4380)$ and the $P_c$s related to the $\Sigma_c^*\bar{D}^*$ threshold,  
which are insignificant in the $\Lambda_b$ decay,
 will complete the spectroscopy in the hadronic molecular picture. 
 That will confirm their $\Sigma_c^{(*)}\bar{D}^{(*)}$ molecular picture. 
Their cross sections are several $\mathrm{nb}$ and one would expect several tens events
based on the current integrated luminosity of LHC at $\sqrt{s}=7~\mathrm{TeV}$. 
The cross sections decrease dramatically with the increasing transverse momentum due to the decreasing behavior of fragmentation functions.
In addition, as these seven hidden charm pentaquarks are  close to the 
corresponding thresholds, there are also sizable deviations of the charged $P_c^+$ cross sections from those of the neutral ones.
The different patterns of the cross sections
could be used to distinguish the two solutions in the molecular picture, which will help us to identify the quantum numbers of the $P_c(4440)$ and $P_c(4457)$.

\begin{acknowledgements}
The discussions with Eulogio Oset, Feng-Kun Guo and Hongxi Xing are appreciated. 
This work is partly supported by Guangdong Major Project of Basic and Applied Basic Research No.~2020B0301030008,
the National Natural Science Foundation of China with Grant No.~12035007, 
Science and Technology Program of Guangzhou No.~2019050001,
Guangdong Provincial funding with Grant No.~2019QN01X172.
Q.W. is also supported by the by the NSFC and the Deutsche Forschungsgemeinschaft (DFG, German
Research Foundation) through the funds provided to the Sino-German Collaborative
Research Center TRR110 “Symmetries and the Emergence of Structure in QCD”
(NSFC Grant No. 12070131001, DFG Project-ID 196253076-TRR 110). The work of M.L.D. is supported by the Spanish Ministerio de Economía y Competitividad (MINECO) and the European Regional
Development Fund (ERDF) under contract FIS2017-84038-C2-1-P, by the EU Horizon
2020 research and innovation programme, STRONG-2020 project, under grant agreement No.~824093,  by Generalitat
Valenciana under contract PROMETEO/2020/023.
\end{acknowledgements}

\appendix
\newpage

\section{The effective couplings of the hidden charm pentaquarks to the relevant channels}
\label{app:couplings}
The closeness of the $P_c(4312)$, $P_c(4440)/P_c(4457)$ to the $\Sigma_c\bar{D}$ and the $\Sigma_c\bar{D}^*$
thresholds, respectively, imply their hadronic molecular picture. Several studies~\cite{Du:2021fmf,Xiao:2020frg,Du:2019pij,Pan:2019skd,Liu:2019zvb,Valderrama:2019chc,Liu:2019tjn} have been implemented
to explore their property in the molecular picture. The complete spectroscopy in the $\Sigma_c^{(*)}\bar{D}^{(*)}$
molecular picture should be three $\frac{1}{2}^-$ pentaquarks, $\frac{3}{2}^-$ pentaquarks and one $\frac{5}{2}^-$ pentaquark.
The heavy quark symmetry constraints that the number of parameters for the underlying dynamics is two,
 i.e. $C_1$ and $C_3$  defined as Eq.(4) of Ref.~\cite{Du:2021fmf}.  The corresponding channels and potentials 
 in each channel are as follow. The three channels for $J^P=\frac{1}{2}^{-}$ are
$\Sigma_{c}\bar{D}$, $\Sigma_{c}\bar{D}^{*}$, $\Sigma_{c}^{*}\bar{D}^{*}$ and the corresponding potential is
\bea
V_{\frac{1}{2}^{-}}=\left(\begin{array}{ccc}
\frac{1}{3}C_{1}+\frac{2}{3}C_{3} & -\frac{2}{3\sqrt{3}}C_{1}+\frac{2}{3\sqrt{3}}C_{3} & \frac{1}{3}\sqrt{\frac{2}{3}}C_{1}-\frac{1}{3}\sqrt{\frac{2}{3}}C_{3}\\
-\frac{2}{3\sqrt{3}}C_{1}+\frac{2}{3\sqrt{3}}C_{3} & \frac{7}{9}C_{1}+\frac{2}{9}C_{3} & -\frac{1}{3}\sqrt{\frac{5}{3}}C_{1}+\frac{1}{3}\sqrt{\frac{5}{3}}C_{3}\\
\frac{1}{3}\sqrt{\frac{2}{3}}C_{1}-\frac{1}{3}\sqrt{\frac{2}{3}}C_{3} & -\frac{1}{3}\sqrt{\frac{5}{3}}C_{1}+\frac{1}{3}\sqrt{\frac{5}{3}}C_{3} & \frac{5}{9}C_{1}+\frac{4}{9}C_{3}
\end{array}\right).
\eea
The case for the $J^P=\frac{3}{2}^{-}$ channel, the dynamical channels are $\Sigma_{c}\bar{D}^{*}$, $\Sigma_{c}^{*}\bar{D}$, $\Sigma_{c}^{*}\bar{D}^{*} $
 and the corresponding potential is
\bea
V_{\frac{3}{2}^{-}}=\left(\begin{array}{ccc}
\frac{1}{9}C_{1}+\frac{8}{9}C_{3} & -\frac{1}{3\sqrt{3}}C_{1}+\frac{1}{3\sqrt{3}}C_{3} & \frac{\sqrt{5}}{9}C_{1}-\frac{\sqrt{5}}{9}C_{3}\\
-\frac{1}{3\sqrt{3}}C_{1}+\frac{1}{3\sqrt{3}}C_{3} & \frac{1}{3}C_{1}+\frac{2}{3}C_{3} & -\frac{1}{3}\sqrt{\frac{5}{3}}C_{1}+\frac{1}{3}\sqrt{\frac{5}{3}}C_{3}\\
\frac{\sqrt{5}}{9}C_{1}-\frac{\sqrt{5}}{9}C_{3} & -\frac{1}{3}\sqrt{\frac{5}{3}}C_{1}+\frac{1}{3}\sqrt{\frac{5}{3}}C_{3} & \frac{5}{9}C_{1}+\frac{4}{9}C_{3}
\end{array}\right).
\eea
There is only one channel, i.e. $\Sigma_{c}^{*}\bar{D}^{*}$ for the $J^P=\frac{5}{2}^{-}$ channel and the potential is
\bea
V_{\frac{5}{2}^{-}}=C_{3}.
\eea

In the whole manuscript, we use the scattering amplitudes of pure contact results in Ref.~\cite{Du:2021fmf} 
as inputs. The inclusion of the OPE and higher order contact potentials
will not change the results significantly. When fit to the $J/\psi p$ invariant mass distribution
of the $\Lambda_b\to J/\psi p K^-$ process, two solutions can be found
~\cite{Du:2021fmf,Du:2019pij}, 
i.e. Solution A and Solution B as denoted in Refs.~\cite{Liu:2019zvb,Liu:2019tjn}.
In Solution A, the $P_c(4440)$ and the $P_c(4457)$ are assigned as $\frac{1}{2}^-$ and $\frac{3}{2}^-$ 
pentaquarks, respectively. In Solution B, they are interchanged. In the whole manuscript,
we use the effective couplings with hard cutoff $1~\gev$ in Ref.~\cite{Du:2021fmf} as inputs. 
In the following, the effective couplings are collected in Tables
~\ref{table:couplingsA1}-\ref{table:couplingsB1}.

\begin{table}
\caption{The effective couplings of each channel for the three $\frac{1}{2}^-$, three $\frac{3}{2}^-$ and one $\frac{5}{2}^-$
pentaquarks in Solution A with hard cutoff $1~\gev$. The second row is the corresponding pole positions. These values are taken from Ref.~\cite{Du:2021fmf}.}
\begin{tabular}{|c|c|c|c|c|}
\hline 
$g_{\text{eff}}(\frac{1}{2}^{-})$ & Pole(MeV) & $\Sigma_{c}\bar{D}$ & $\Sigma_{c}\bar{D}^{*}$ & $\Sigma_{c}^{*}\bar{D}^{*}$\tabularnewline
\hline 
\hline 
$P_{c}(4312)$ & $4314.7-i3.4$ & $2.63+i0.33$ & $0.77+i0.1$ & $0.44+i0.05$\tabularnewline
\hline 
$P_{c}(4440)$ & $4439.8-i7.2$ & $0.15+i0.35$ & $3.74+i0.46$ & $-0.75+i0.3$\tabularnewline
\hline 
$P_{c}(\frac{1}{2}^{-})$ & $4497.4-i6.9$ & $0.21+i0.16$ & $-0.06-i0.27$ & $4.09+i0.24$\tabularnewline
\hline 
\hline 
$g_{\text{eff}}(\frac{3}{2}^{-})$ & Pole(MeV) & $\Sigma_{c}\bar{D}^{*}$ & $\Sigma_{c}^{*}\bar{D}$ & $\Sigma_{c}^{*}\bar{D}^{*}$\tabularnewline
\hline 
\hline 
$P_{c}(4380)$ & $4377.2-i6.1$ & $0.50+i0.25$ & $-2.75-i0.05$ & $-0.95+i0.16$\tabularnewline
\hline 
$P_{c}(4457)$ & $4459.3-i3.0$ & $2.06+i0.44$ & $0.05-i0.09$ & $-0.78+i0.32$\tabularnewline
\hline 
$P_{c}(\frac{3}{2}^{-})$ & $4506.9-i18.6$ & $0.37+i0.32$ & $-0.26-i0.35$ & $-3.63-i0.85$\tabularnewline
\hline 
\hline 
\hline 
$g_{\text{eff}}(\frac{5}{2}^{-})$ & Pole(MeV) & $\Sigma_{c}^{*}\bar{D}^{*}$&&\tabularnewline
\hline 
\hline 
$P_{c}(\frac{5}{2}^{-})$ & $4526.7-i13.6$ & $2.18+i1.29$&&\tabularnewline
\hline 
\end{tabular}
\label{table:couplingsA1}
\end{table}

\begin{table}
\caption{The caption is analogous to that of Table~\ref{table:couplingsA1} but for Solution B. These values are taken from Ref.~\cite{Du:2021fmf}.}
\begin{tabular}{|c|c|c|c|c|}
\hline 
$g_{\text{eff}}(\frac{1}{2}^{-})$ & Pole(MeV) & $\Sigma_{c}\bar{D}$ & $\Sigma_{c}\bar{D}^{*}$ & $\Sigma_{c}^{*}\bar{D}^{*}$\tabularnewline
\hline 
\hline 
$P_{c}(4312)$ & $4312.0-i4.8$ & $2.92+i0.43$ & $-0.77+i0.07$ & $-0.51+i0.08$\tabularnewline
\hline 
$P_{c}(4457)$ & $4466.9-i7.1$ & $0.08+i0.41$ & $-2.02-i1.58$ & $-0.32-i0.28$\tabularnewline
\hline 
$P_{c}(\frac{1}{2}^{-})$ & $4530.5-i9.7$ & $0.03+i0.20$ & $0.04-i0.18$ & $-1.59-i1.67$\tabularnewline
\hline 
\hline 
$g_{\text{eff}}(\frac{3}{2}^{-})$ & Pole(MeV) & $\Sigma_{c}\bar{D}^{*}$ & $\Sigma_{c}^{*}\bar{D}$ & $\Sigma_{c}^{*}\bar{D}^{*}$\tabularnewline
\hline 
\hline 
$P_{c}(4380)$ & $4374.0-i5.9$ & $0.58-i0.20$ & $3.05+i0.10$ & $-0.87+i0.18$\tabularnewline
\hline 
$P_{c}(4440)$ & $4441.6-i4.6$ & $3.63+i0.32$ & $0.03+i0.08$ & $0.84-i0.14$\tabularnewline
\hline 
$P_{c}(\frac{3}{2}^{-})$ & $4522.2-i15.1$ & $0.00-i0.31$ & $0.20+i0.38$ & $-2.51-i1.10$\tabularnewline
\hline 
\hline 
$g_{\text{eff}}(\frac{5}{2}^{-})$ & Pole(MeV) & $\Sigma_{c}^{*}\bar{D}^{*}$&&\tabularnewline
\hline 
\hline 
$P_{c}(\frac{5}{2}^{-})$ & $4500.5-i4.3$ & $3.95+i0.07$&&\tabularnewline
\hline 
\end{tabular}
\label{table:couplingsB1}
\end{table}

\section{Cross section of the inclusive production of the $\Sigma_c^{(*)}\bar{D}^{(*)}$ pair in the LHC}
\label{app:SigmacDbar}
\begin{figure}[htbp]
\begin{center}
 \includegraphics[width=0.55\textwidth]{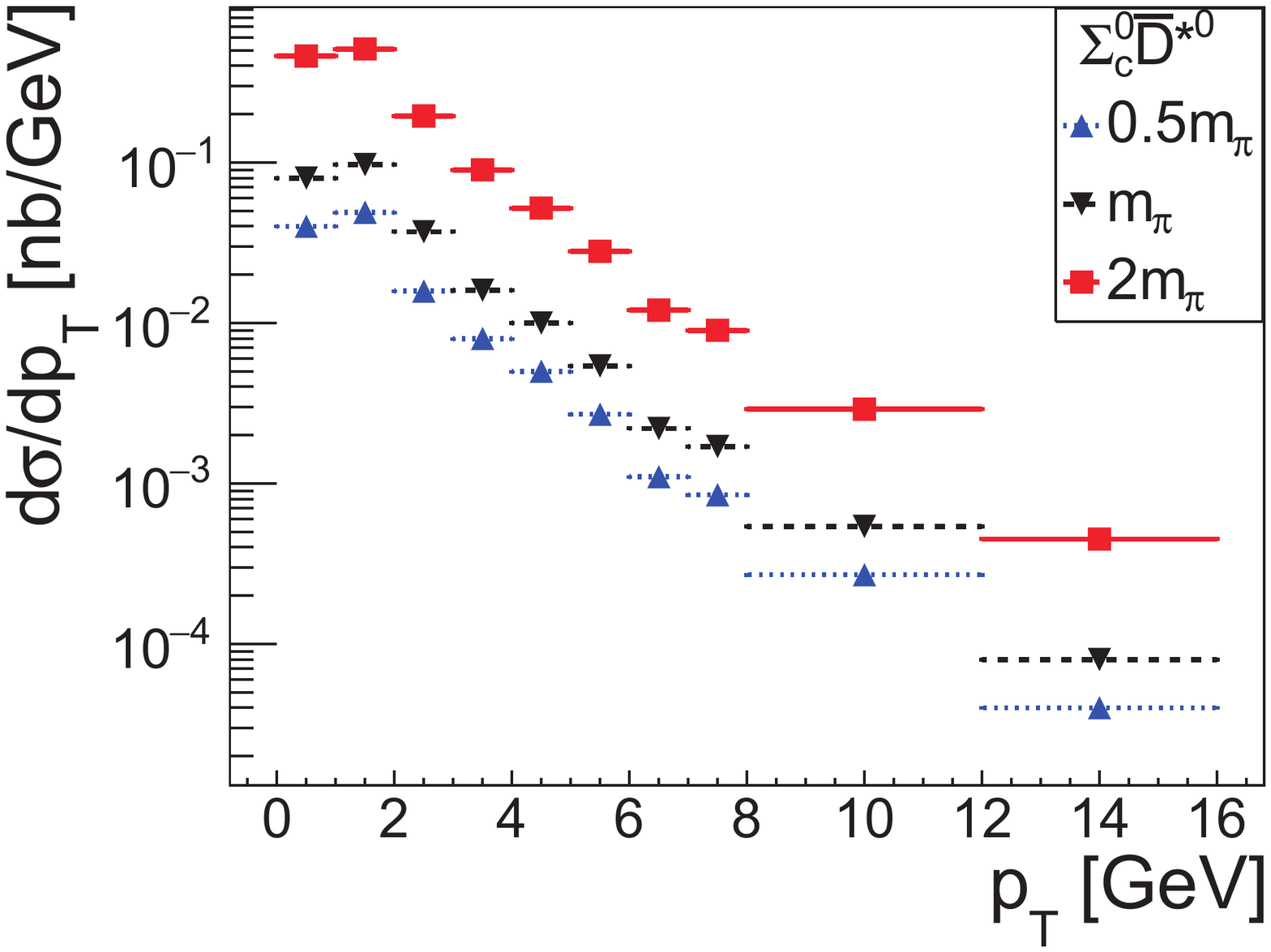} \hspace{-2cm}
 \includegraphics[width=0.55\textwidth]{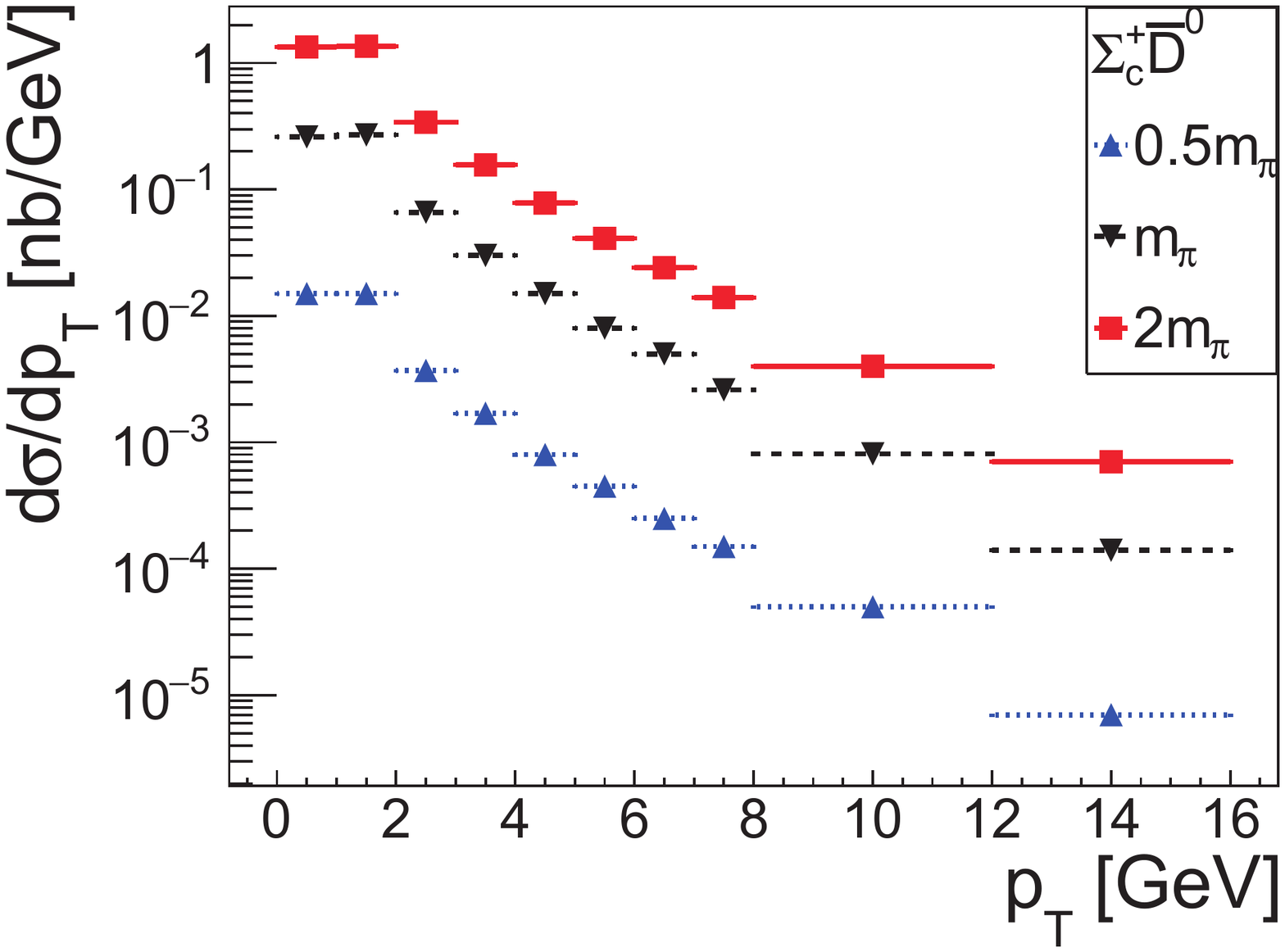} \\
 \includegraphics[width=0.55\textwidth]{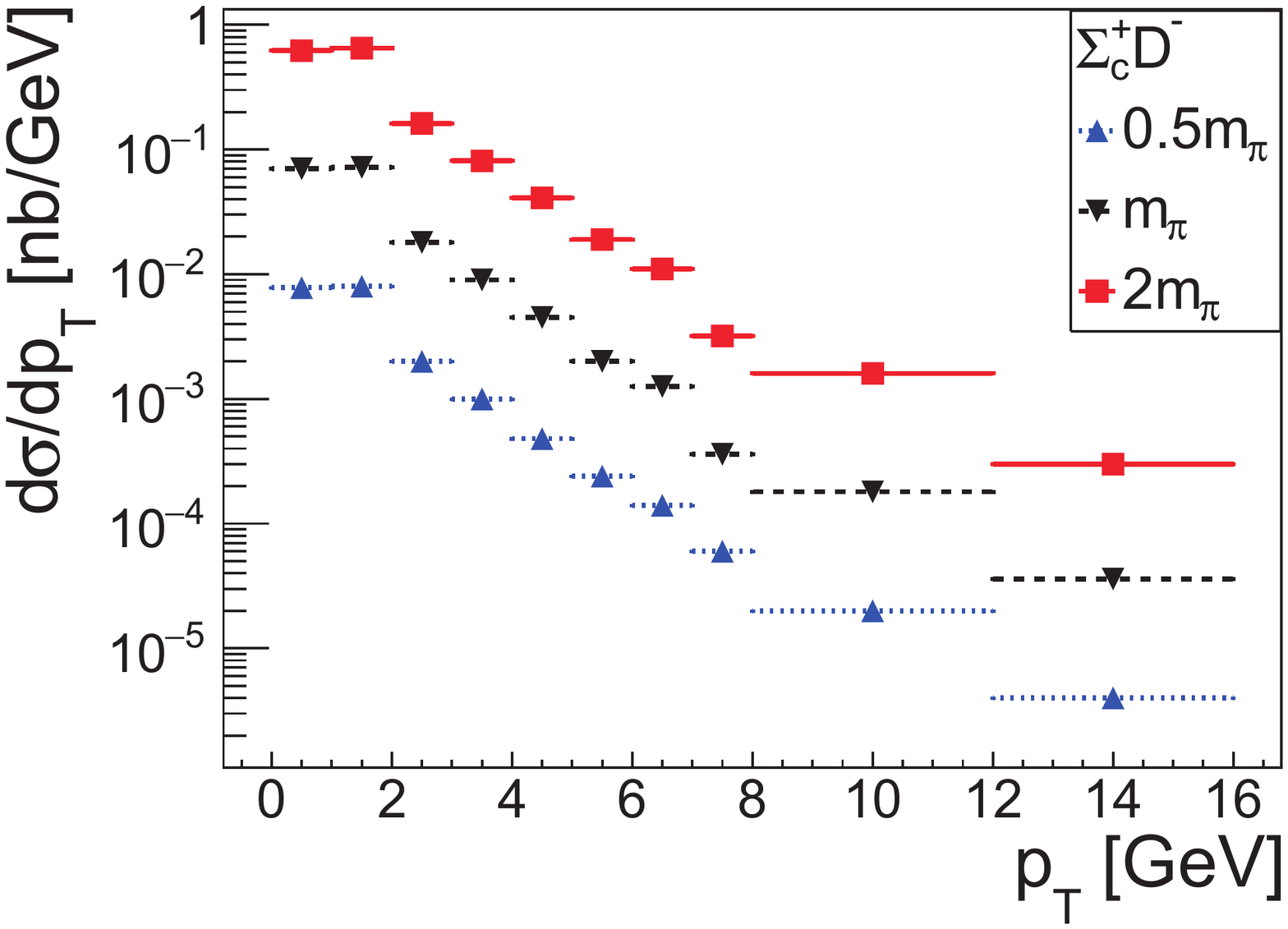} \hspace{-2cm}
 \includegraphics[width=0.55\textwidth]{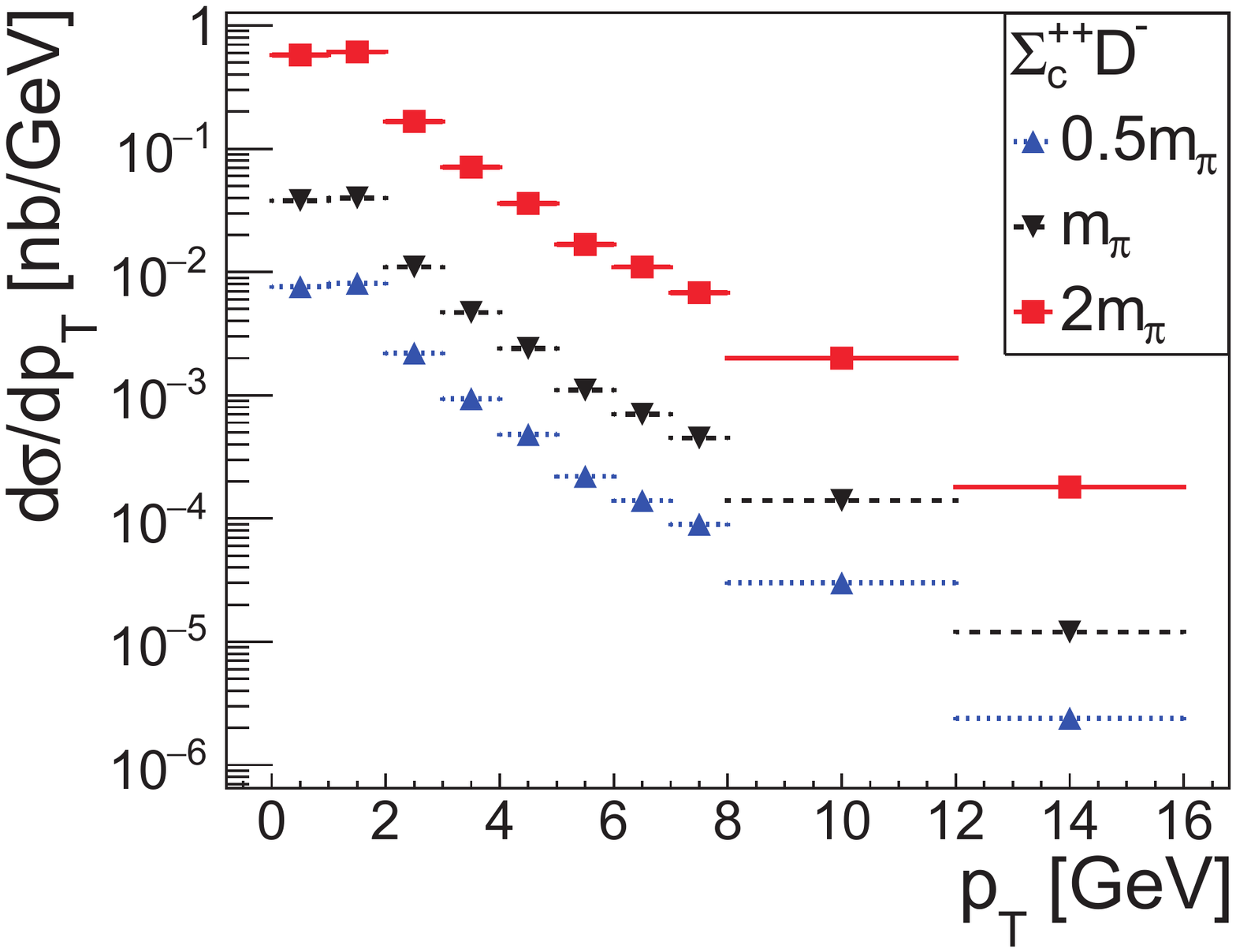}\\
\caption{The cross section of the inclusive production of the $\Sigma_c\bar{D}$ pairs in $pp$ collisions 
at center-of-mass energy $\sqrt{s}=7~\tev$. The    
black inverted triangle, blue triangle, red box points are for the relative three momentum
smaller than $0.5m_\pi$, $m_\pi$ and $2m_\pi$, respectively. As shown in the figures,
there are significant isospin deviation for the inclusive productions of the $\Sigma_c\bar{D}$ pairs.
In our calculation, we use $m_\pi$ to estimate the prompt cross section.}
\label{fig:SigmacDbar}
\end{center}
\end{figure}

\begin{figure}[htbp]
\begin{center}
 \includegraphics[width=0.55\textwidth]{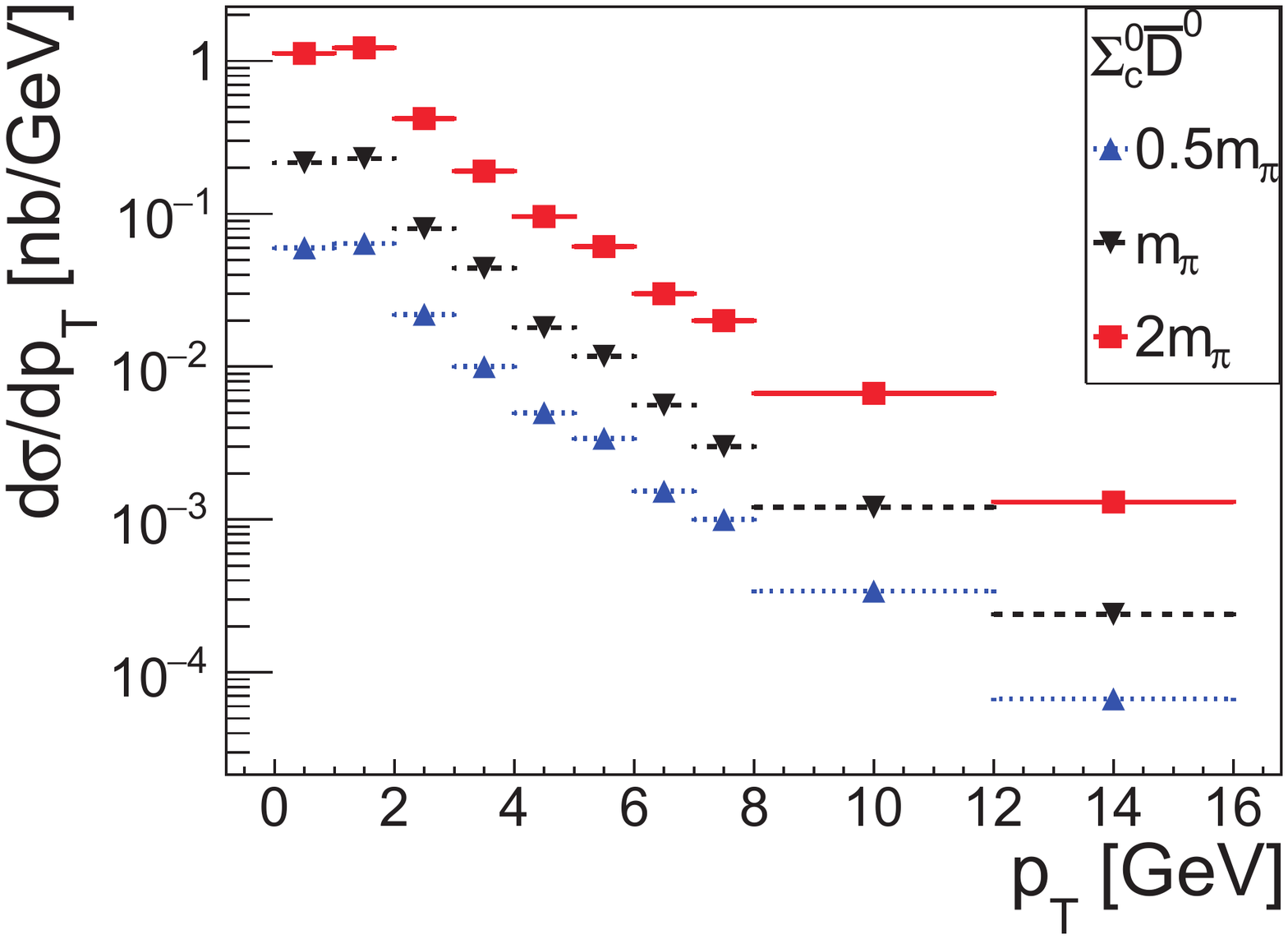} \hspace{-2cm}
 \includegraphics[width=0.55\textwidth]{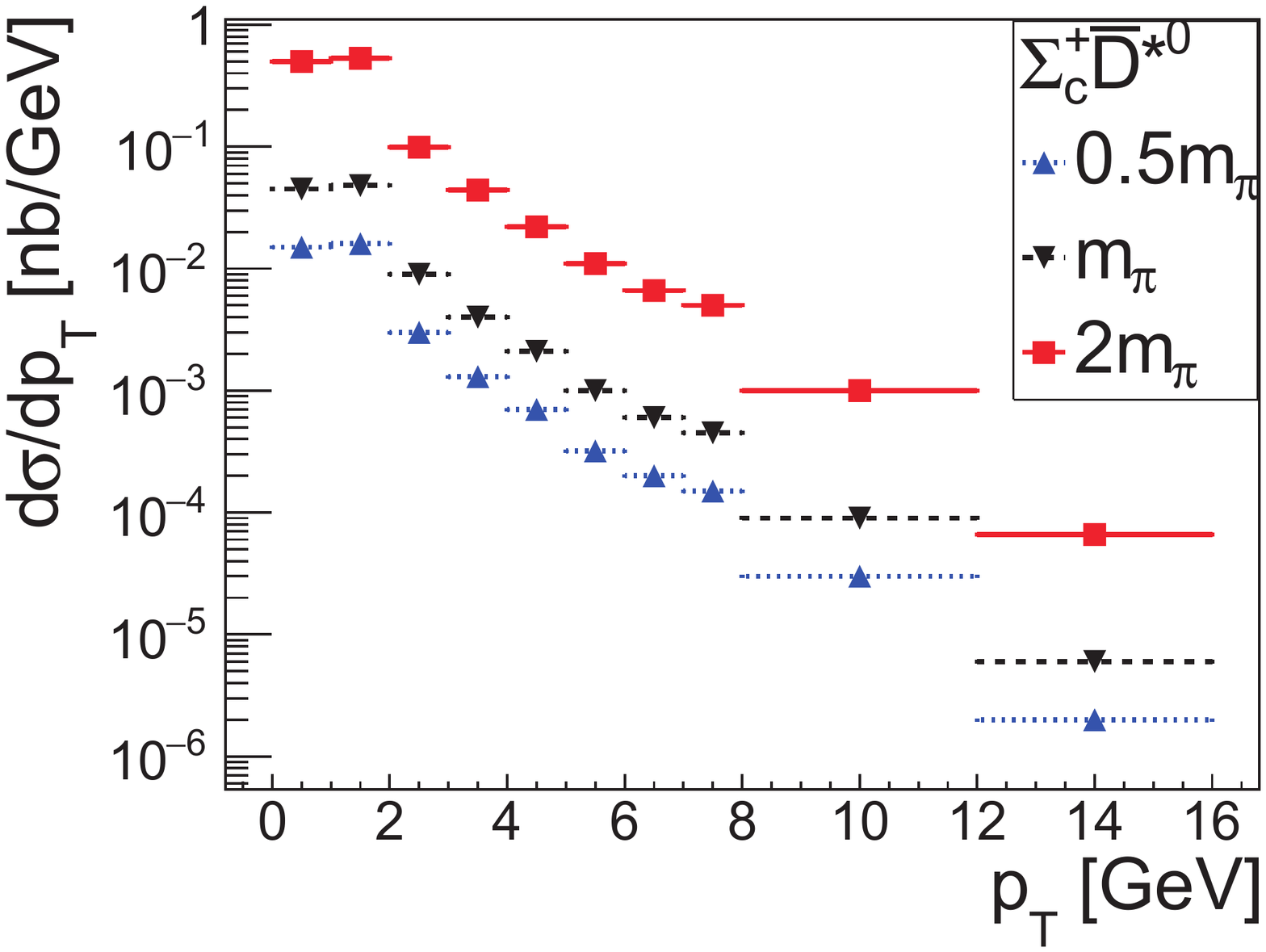} \\
 \includegraphics[width=0.55\textwidth]{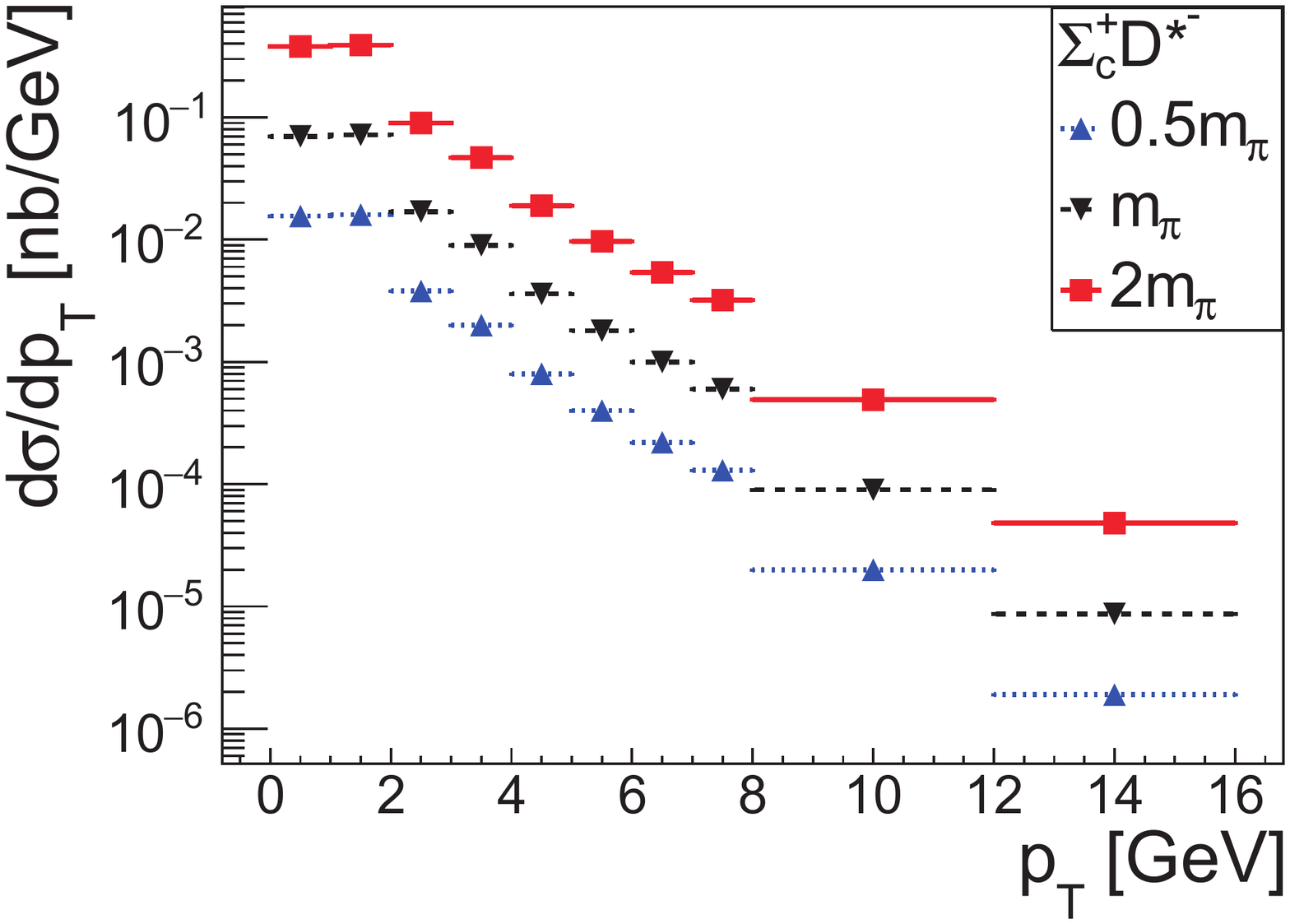} \hspace{-2cm}
 \includegraphics[width=0.55\textwidth]{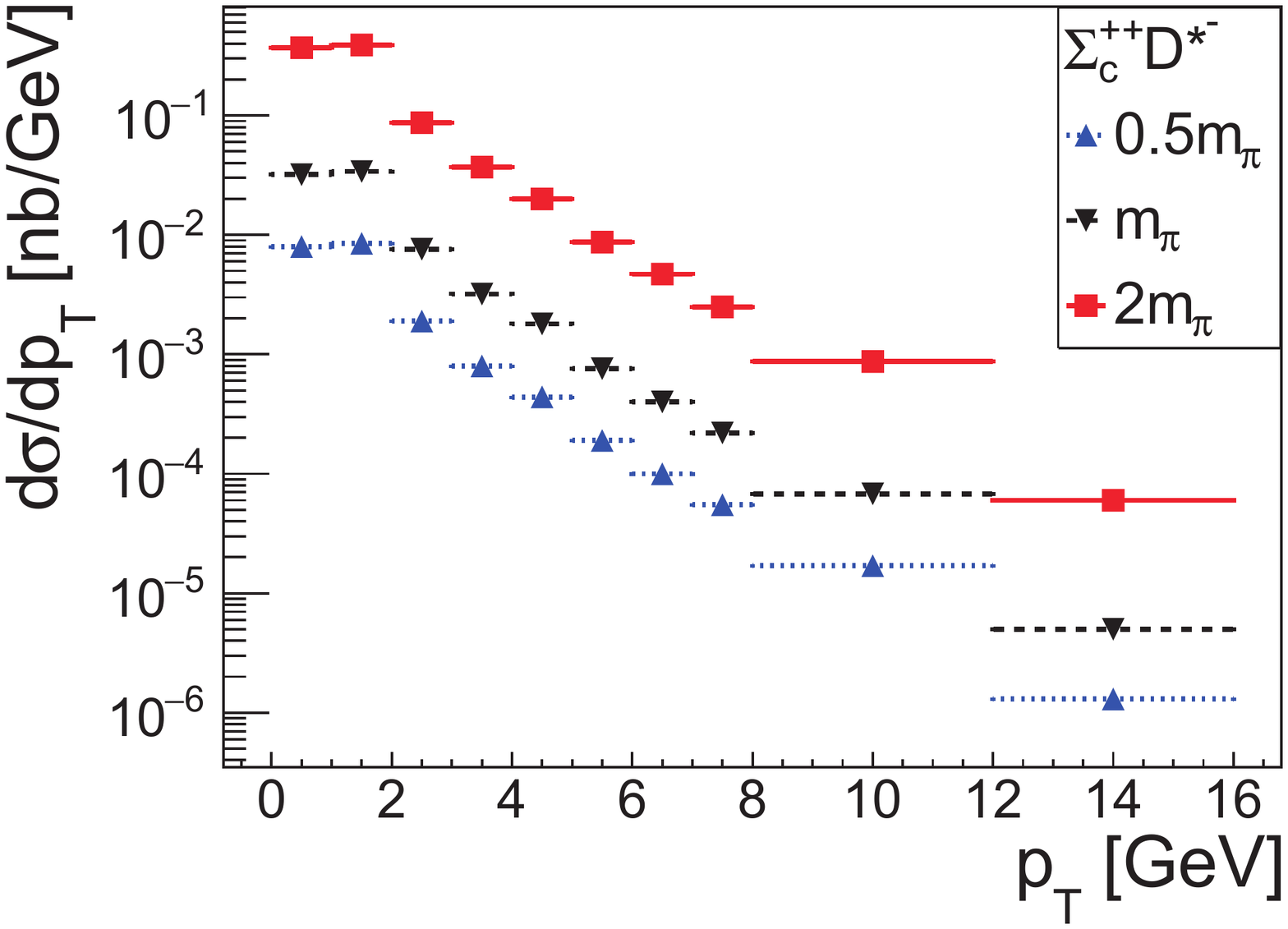}\\
\caption{The caption is analogous to that of Fig.~\ref{fig:SigmacDbar}, but for the $\Sigma_c\bar{D}^*$ channel.}
\label{fig:SigmacDstarbar}
\end{center}
\end{figure}

\begin{figure}[htbp]
\begin{center}
 \includegraphics[width=0.55\textwidth]{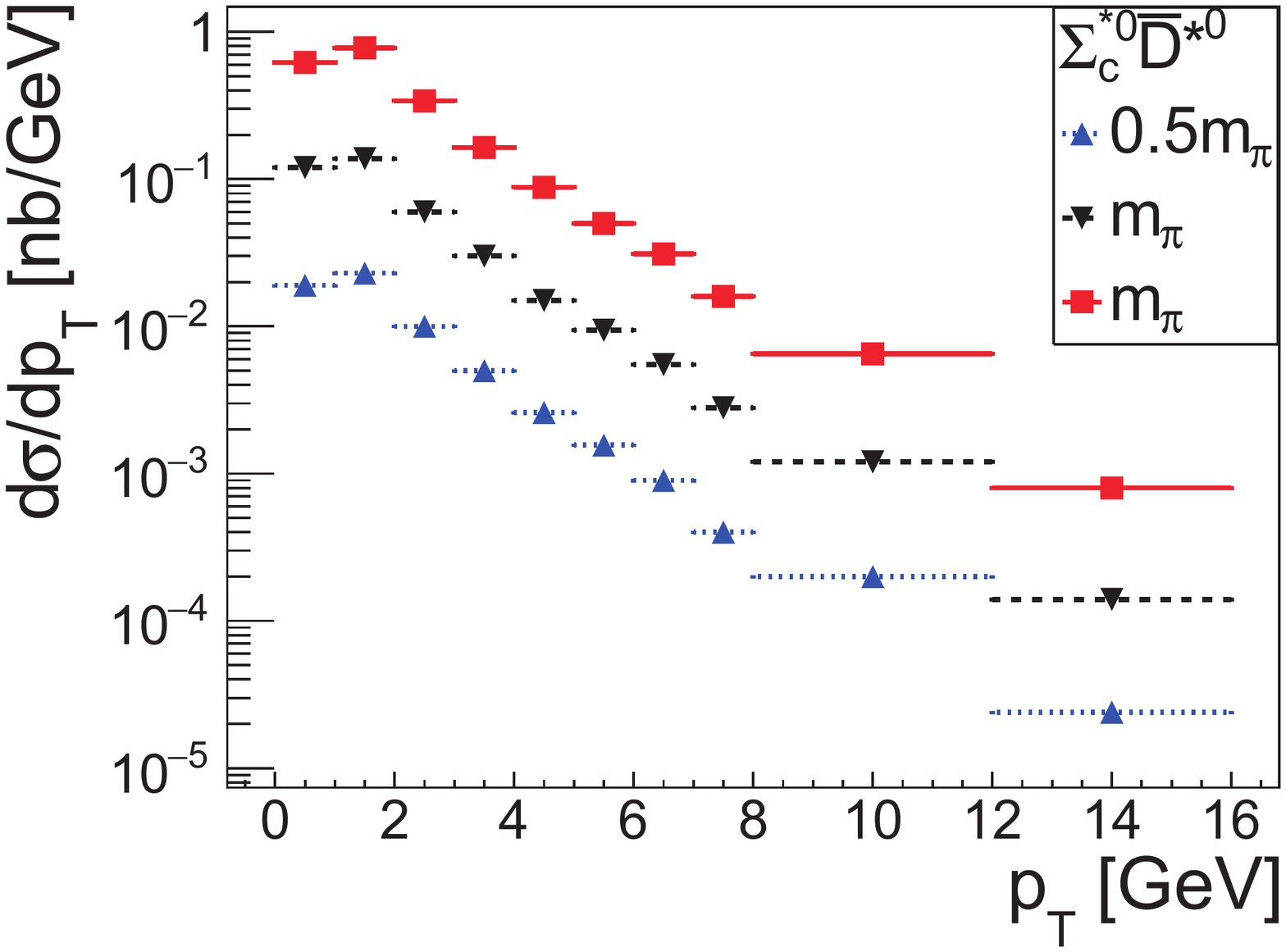} \hspace{-2cm}
 \includegraphics[width=0.55\textwidth]{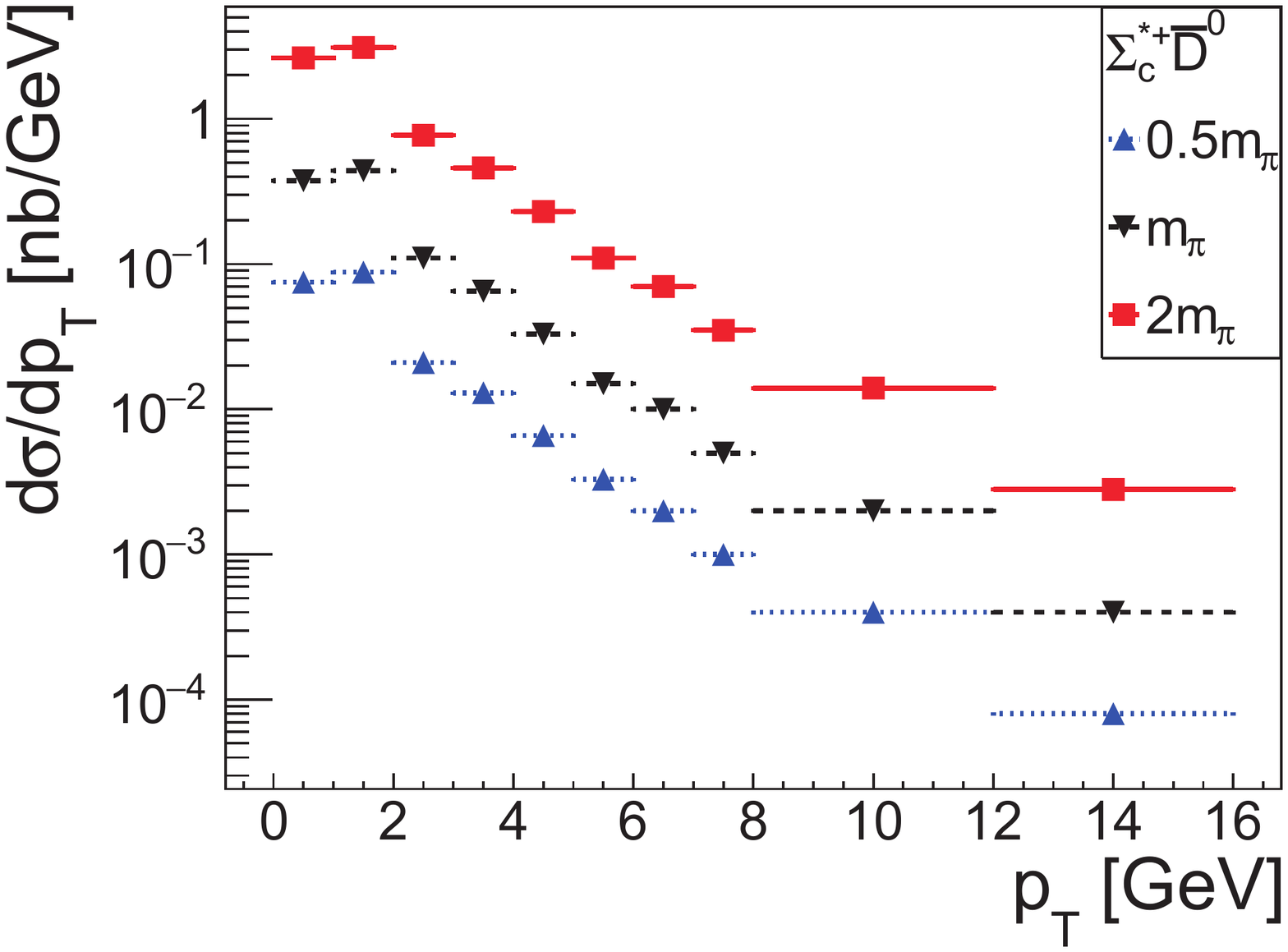} \\
 \includegraphics[width=0.55\textwidth]{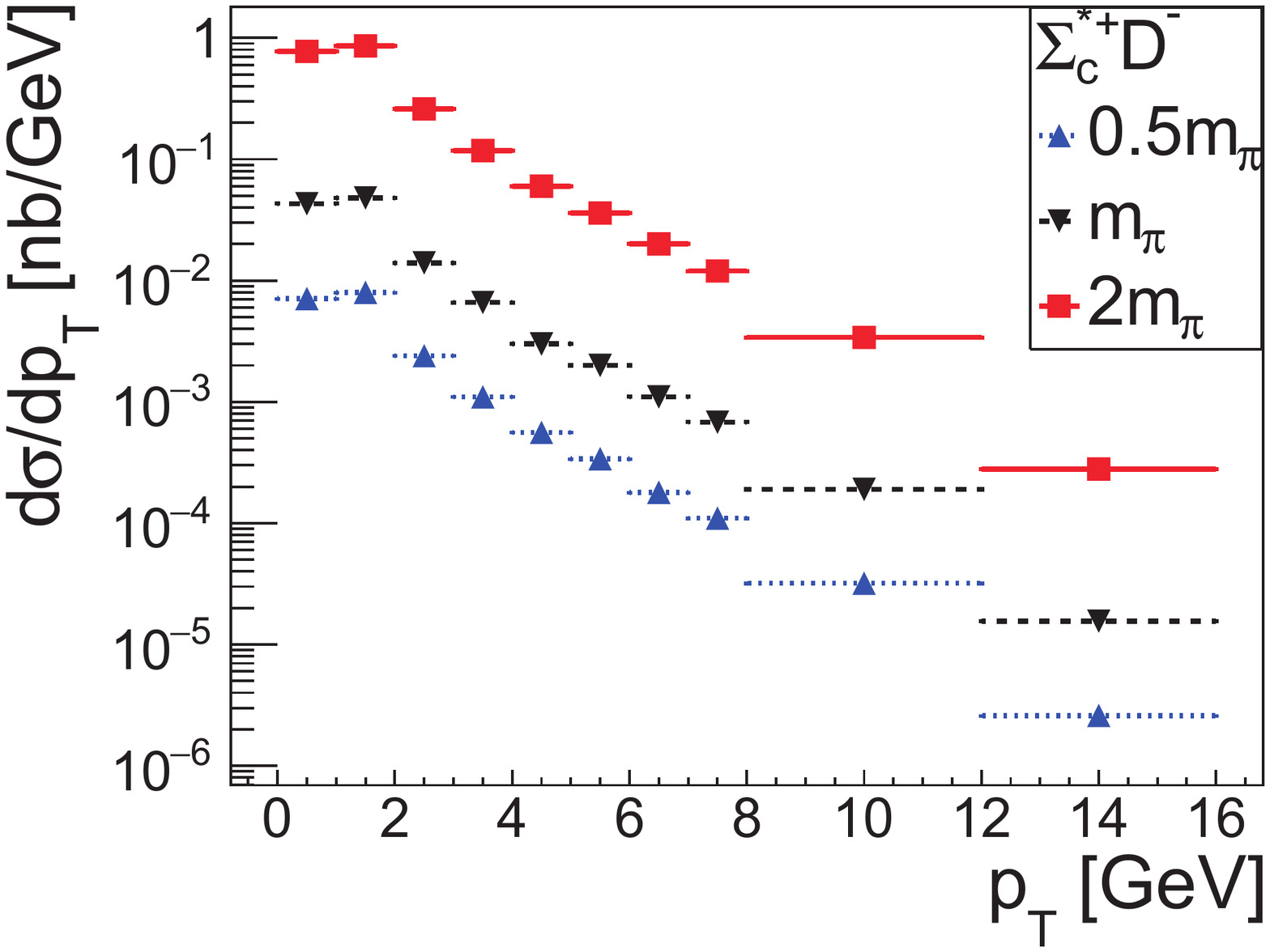} \hspace{-2cm}
 \includegraphics[width=0.55\textwidth]{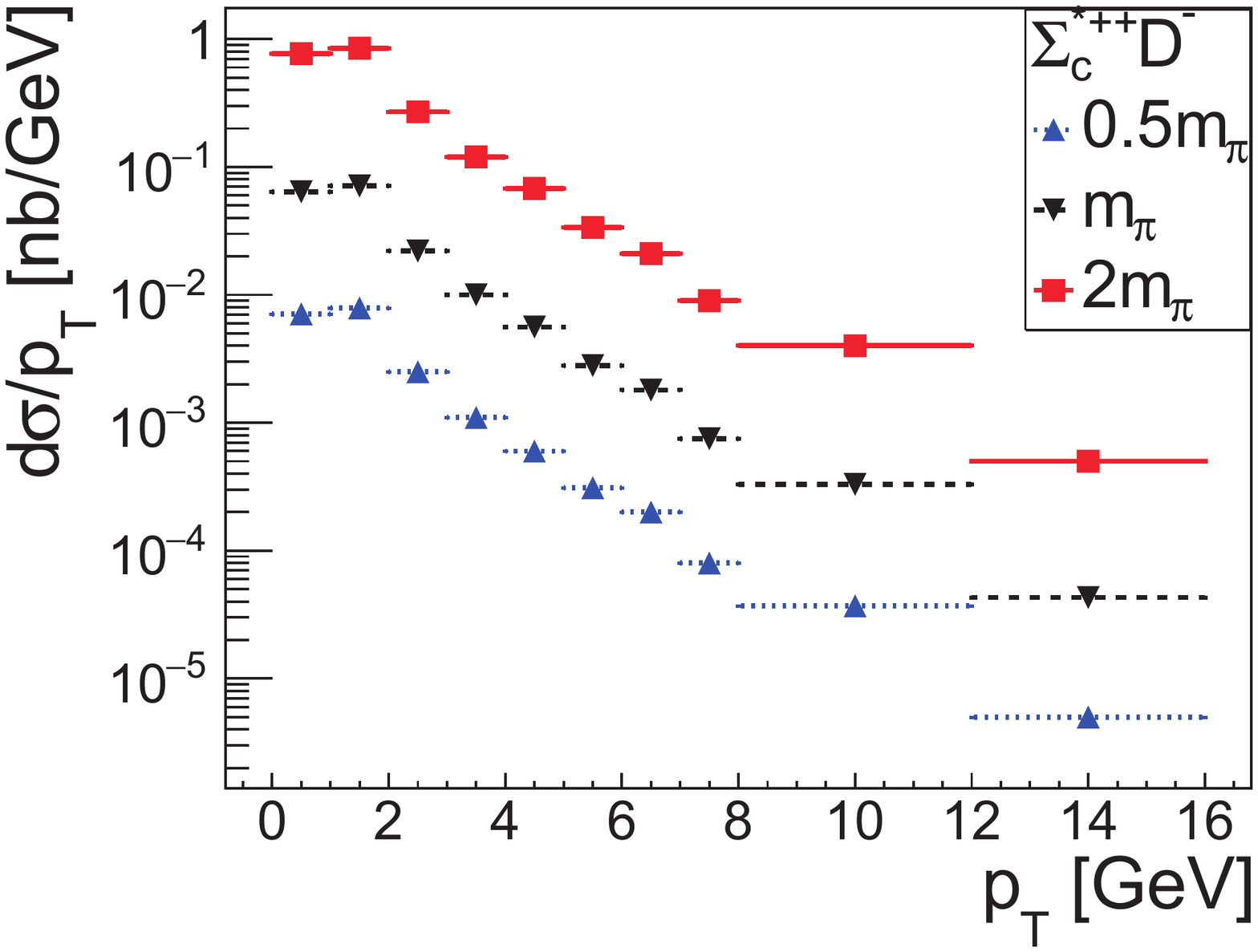}\\
\caption{The caption is analogous to that of Fig.~\ref{fig:SigmacDbar}, but for the $\Sigma_c^*\bar{D}$ channel.}
\label{fig:SigmacstarDbar}
\end{center}
\end{figure}

\begin{figure}[htbp]
\begin{center}
 \includegraphics[width=0.55\textwidth]{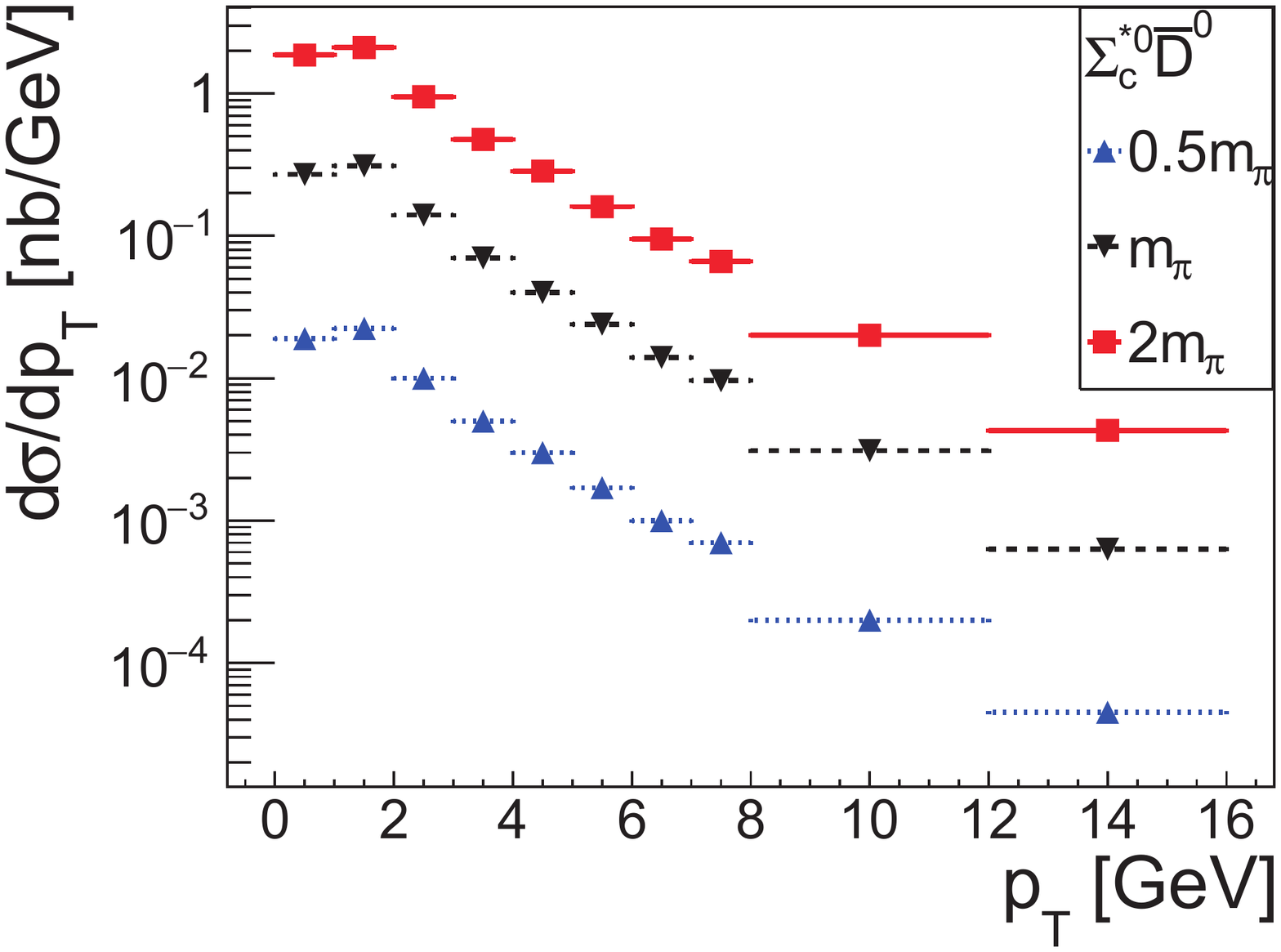} \hspace{-2cm}
 \includegraphics[width=0.55\textwidth]{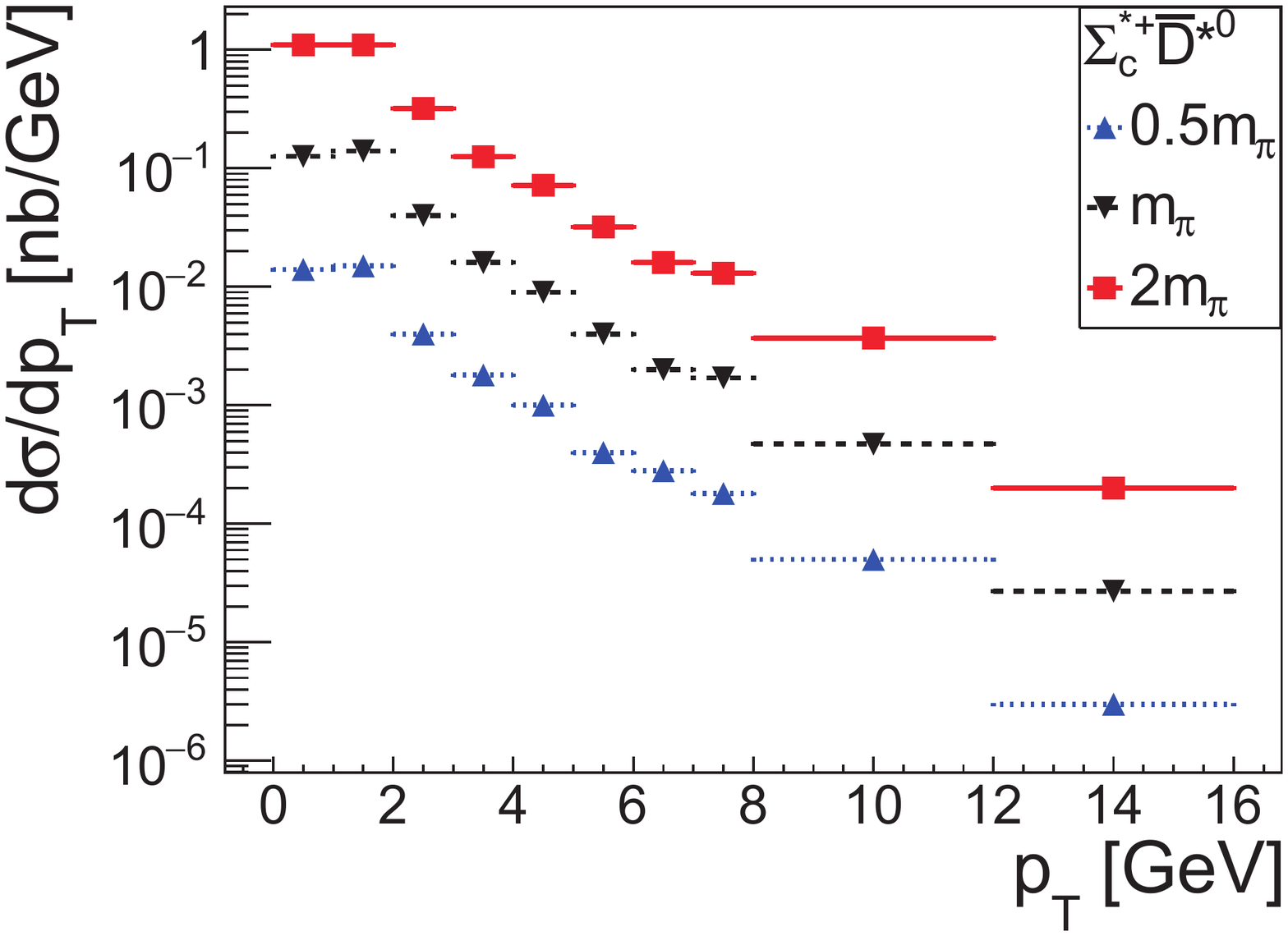} \\
 \includegraphics[width=0.55\textwidth]{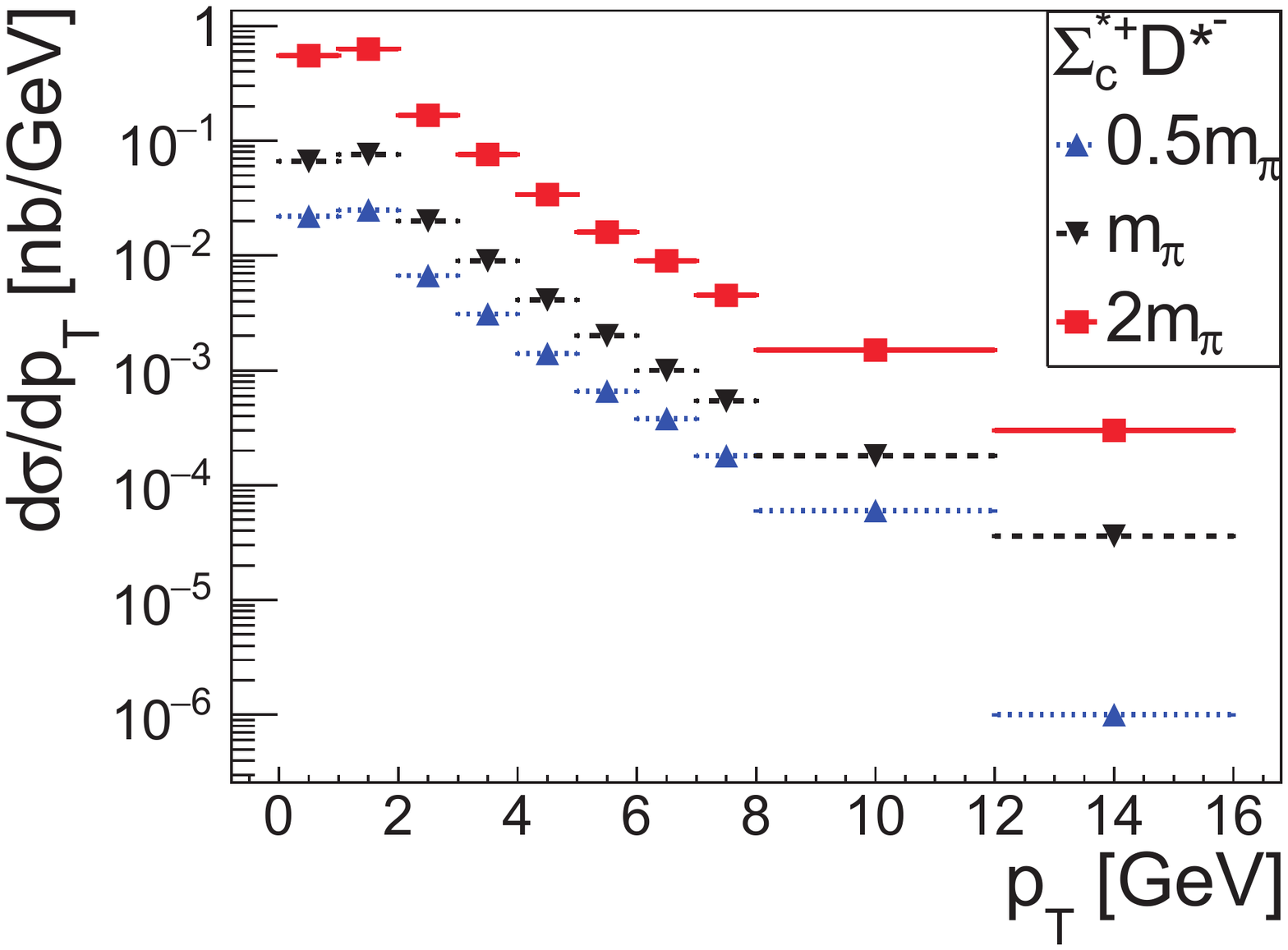} \hspace{-2cm}
 \includegraphics[width=0.55\textwidth]{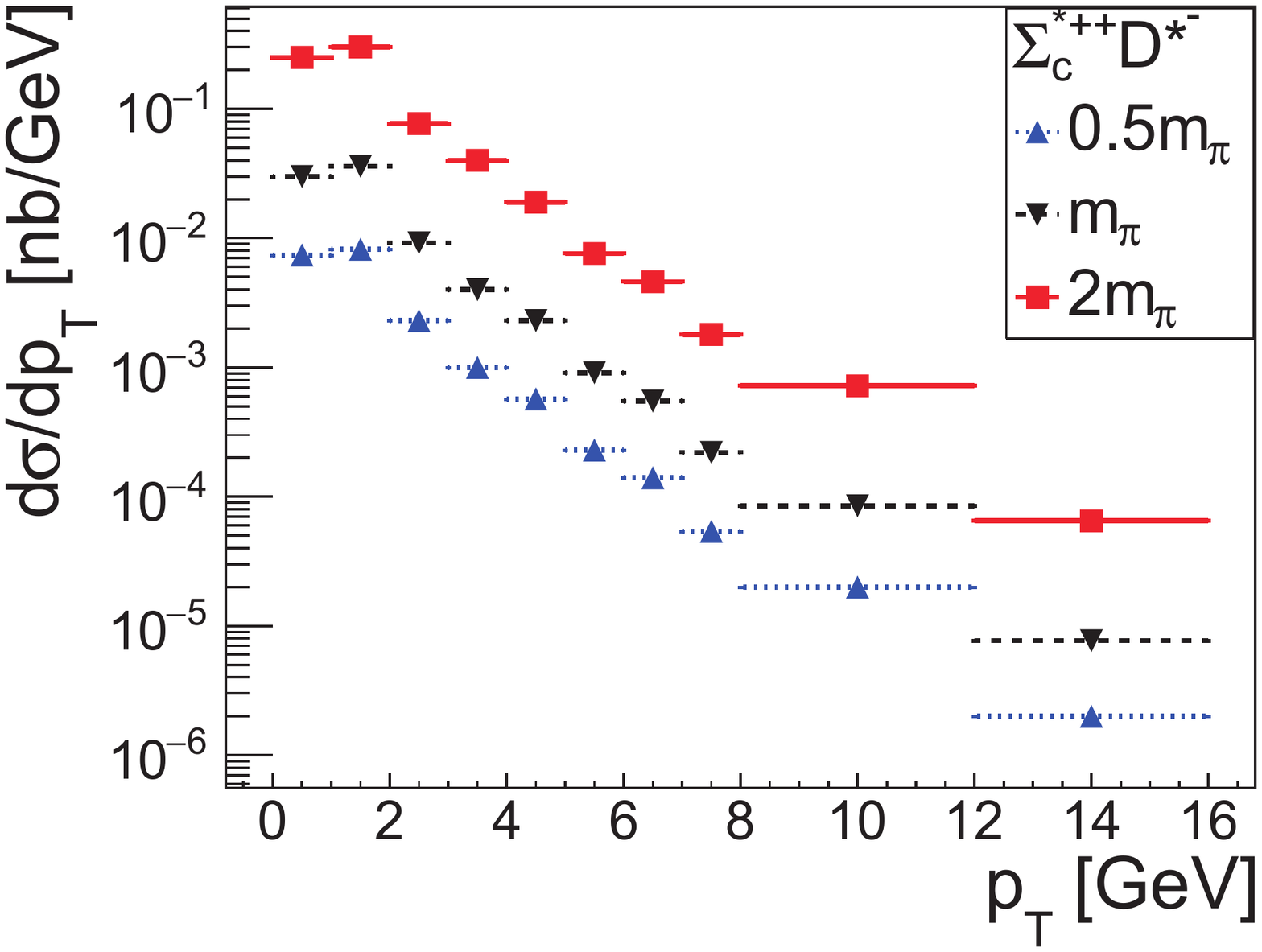}\\
\caption{The caption is analogous to that of Fig.~\ref{fig:SigmacDbar}, but for the $\Sigma_c^*\bar{D}^*$ channel.}
\label{fig:SigmacstarDstarbar}
\end{center}
\end{figure}

The property that the seven hidden charm pentaquarks, in the molecular picture, are very close
the $\Sigma_c^{(*)}\bar{D}^{(*)}$ thresholds does not only lead to a large isospin breaking decay rate 
(such as the $J/\psi\Delta^+$ channel~\cite{Guo:2019fdo}), but also could affect their production rates.  That can be easily 
seen from Eq.\eqref{eq:two-body}, i.e. a small deviation of the thresholds will result in a large difference
in the two-body propagator. If isospin symmetry works well, the inclusive production cross section of the
$\Sigma_c^{(*)+}\bar{D}^{(*)0}$ ($\Sigma_c^{(*)+}D^{(*)-}$) channel should equal to that of the
 $\Sigma_c^{(*)++}D^{(*)-}$ ($\Sigma_c^{(*)0}\bar{D}^{(*)0}$) channel. However, the initial $pp$ beam has 
 a third isospin $I_z=+1$ which indicates that one would not expect the inclusive 
 $I_z=+\frac{1}{2}$ $\Sigma_c^{(*)+}\bar{D}^{(*)0}$, $\Sigma_c^{(*)++}D^{(*)-}$ channels 
 have the same inclusive cross section as those of the  $I_z=-\frac{1}{2}$ $\Sigma_c^{(*)+}D^{(*)-}$, $\Sigma_c^{(*)0}\bar{D}^{(*)0}$
 channels. Another reason is that we do not know the isospin information of the undetected particles for the inclusive processes.
 To illustrate this effect, we plot the cross sections of the inclusive $\Sigma_c^{(*)}\bar{D}^{(*)}$ channels explicitly
 in Figs.~\ref{fig:SigmacDbar}-\ref{fig:SigmacstarDstarbar}. From the figures, one can also see the 
 relation of the production rates for all the $\Sigma_c^{(*)}\bar{D}^{(*)}$ channels
 \bea
\sigma\left(\Sigma_{c}^{*}\bar{D}\right)>\sigma\left(\Sigma_{c}\bar{D}\right)>\sigma\left(\Sigma_{c}\bar{D}^{*}\right)\backsimeq\sigma\left(\Sigma_{c}^{*}\bar{D}^{*}\right).
\label{eq:relation_of_cs}
\eea

\end{document}